%% file: GW-MMADS_S230922g.tex
\documentclass[twocolumn]{aastex631}

\usepackage[acronym]{glossaries}
\usepackage{xspace}
\usepackage{ulem}

\newcommand{\gweventid}{S230922g\xspace}
\newcommand{\gwemopt}{\texttt{gwemopt}\xspace}

\newcommand{\mcwilliams}{
    McWilliams Center for Cosmology and Astrophysics,
    Department of Physics,
    Carnegie Mellon University,
    5000 Forbes Avenue, Pittsburgh, PA 15213
}

\newcommand{\usm}{
    University Observatory,
    Faculty of Physics,
    Ludwig-Maximilians-Universität München,
    Scheinerstr. 1, 81679 Munich, Germany
}
\definecolor{maroon}{rgb}{0.5, 0.0, 0.0}
\newcommand{\response}[1]{#1\xspace}

\newacronym{gw}{GW}{gravitational wave}
\newacronym{grb}{GRB}{gamma ray burst}
\newacronym{ns}{NS}{neutron star}
\newacronym{bh}{BH}{black hole}
\newacronym{bns}{BNS}{binary neutron star}
\newacronym{bbh}{BBH}{binary black hole}
\newacronym{em}{EM}{electromagnetic}
\newacronym[plural=AGNs, firstplural=active galactic nuclei]{agn}{AGN}{active galactic nucleus}
\newacronym{sfft}{\texttt{SFFT}}{\texttt{Saccadic Fast Fourier Transform}}
\newacronym{lvk}{LVK}{LIGO/Virgo/KAGRA}
\newacronym{o4}{O4}{the fourth gravitational wave observing run}
\newacronym{o5}{O5}{the fifth gravitational wave observing run}
\newacronym{psc}{PSC}{Pittsburgh Supercomputing Center}
\newacronym{nersc}{NERSC}{National Energy Research Scientific Computing Center}
\newacronym{decam}{DECam}{Dark Energy Camera}
\newacronym{cp}{CP}{DECam Community Pipeline}
\newacronym{des}{DES}{Dark Energy Survey}
\newacronym{desi}{DESI}{Dark Energy Spectroscopic Instrument}
\newacronym{decals}{DECaLS}{DECam Legacy Survey}
\newacronym{delve}{DELVE}{DECam Local Volume Exploration survey}
\newacronym{ls}{LS}{DESI Legacy Survey}
\newacronym{mpc}{MPC}{Minor Planet Center}
\newacronym{tns}{TNS}{Transient Name Server}
\newacronym{cbc}{CBC}{compact binary coalescence}
\newacronym{gcn}{GCN}{General Coordinates Network}
\newacronym[plural=KNe, firstplural=kilonovae]{kn}{KN}{kilonova}
\newacronym{too}{ToO}{Target of Opportunity}
\newacronym{healpix}{HEALPix}{Hierarchical Equal Area isoLatitude Pixelation}
\newacronym{gwmmads}{GW-MMADS}{Gravitational Wave MultiMessenger Astronomy DECam Survey}
\newacronym{snr}{SNR}{signal-to-noise ratio}
\newacronym{noirlab}{NOIRLab}{NSF National Optical-Infrared Research Laboratory}
\newacronym{cnn}{CNN}{convolutional neural network}
\newacronym{parsnip}{ParSNIP}{Parameterization of SuperNova Intrinsic Properties}
\newacronym[plural=SNe, firstplural=supernovae]{sn}{SN}{supernova}
\newacronym{blr}{BLR}{broad-line region}
\newacronym{wise}{WISE}{Wide-field Infrared Survey Explorer}
\newacronym{sf}{SF}{structure function}
\newacronym{ned}{NED}{NASA/IPAC Extragalactic Database}
\newacronym{tde}{TDE}{tidal distruption event}
\newacronym{bpt}{BPT}{Baldwin–Phillips–Terlevich}
\newacronym{lsst}{LSST}{Legacy Survey of Space and Time}
\newacronym{ztf}{ZTF}{Zwicky Transient Facility}
\newacronym[plural=SLSNe, firstplural=superluminous supernovae]{slsn}{SLSN}{superluminous supernova}
\newacronym{bhl}{BHL}{Bondi-Hoyle-Littleton}
\newacronym{smbh}{SMBH}{supermassive black hole}
\newacronym{ctio}{CTIO}{Cerro Tololo Inter-American Observatory}
\newacronym{ir}{IR}{infrared}
\newacronym{ci}{CI}{confidence interval}
\newacronym{lris}{LRIS}{Low-Resolution Imaging Spectrograph}
\newacronym{gmos}{GMOS}{Gemini Multi-Object Spectrograph}
\newacronym{dragons}{DRAGONS}{Data Reduction for Astronomy from Gemini Observatory North and South}
\newacronym{salt}{SALT}{South African Large Telescope}
\newacronym{rss}{RSS}{Robert Stobie Spectrograph}
\newacronym{pc}{PC}{Photon Counting}
\newacronym{3kk}{3KK}{Three Channel Camera}
\newacronym{ps1}{PS1}{PanSTARRS 1}
\newacronym{2mass}{2MASS}{Two Micron All Sky Survey}
\newacronym{nir}{NIR}{near-infrared}

\shortauthors{Cabrera et al.}

\begin{document}

\title{Searching for electromagnetic emission in an AGN from the gravitational wave binary black hole merger candidate S230922g\footnote{based on observations made with the Southern African Large Telescope (SALT)}\footnote{Some of the data presented herein were obtained at Keck Observatory, which is a private 501(c)3 non-profit organization operated as a scientific partnership among the California Institute of Technology, the University of California, and the National Aeronautics and Space Administration. The Observatory was made possible by the generous financial support of the W. M. Keck Foundation. }}

\shorttitle{GW-MMADS S230922g Follow-up}

\author[0000-0002-1270-7666]{Tom\'as Cabrera}
\affiliation{\mcwilliams}

\correspondingauthor{Tom\'as Cabrera}
\email{tcabrera@andrew.cmu.edu}

\author[0000-0002-6011-0530]{Antonella Palmese}
\affiliation{\mcwilliams}

\author[0000-0001-7201-1938]{Lei Hu}
\affiliation{\mcwilliams}

\author[0000-0002-9700-0036]{Brendan O'Connor}
\altaffiliation{McWilliams Fellow}
\affiliation{\mcwilliams}

\author[0000-0002-5956-851X]{K.E.Saavik Ford}
\affiliation{Center for Computational Astrophysics, Flatiron Institute, 
162 5th Ave, New York, NY 10010, USA}
\affiliation{Department of Astrophysics, American Museum of Natural History, New York, NY 10024, USA}
\affiliation{Department of Science, BMCC, City University of New York, New York, NY 10007, USA}

\author[0000-0002-0786-7307]{Barry McKernan}
\affiliation{Center for Computational Astrophysics, Flatiron Institute, 
162 5th Ave, New York, NY 10010, USA}
\affiliation{Department of Astrophysics, American Museum of Natural History, New York, NY 10024, USA}
\affiliation{Department of Science, BMCC, City University of New York, New York, NY 10007, USA}

\author[0000-0002-8977-1498]{Igor Andreoni}
\affiliation{Joint Space-Science Institute, University of Maryland, College Park, MD 20742, USA}
\affiliation{Department of Astronomy, University of Maryland, College Park, MD 20742, USA}
\affiliation{Astrophysics Science Division, NASA Goddard Space Flight Center, Mail Code 661, Greenbelt, MD 20771, USA}

\author[0000-0002-2184-6430]{Tom\'as Ahumada}
\affiliation{Division of Physics, Mathematics, and Astronomy, California Institute of Technology, Pasadena, CA 91125, USA}

\author[0000-0003-3433-2698]{Ariel Amsellem}
\affiliation{\mcwilliams}

\author[0009-0001-0574-2332]{Malte Busmann}
\affiliation{\usm}

\author[0000-0002-6576-7400]{Peter Clark}
\affiliation{Institute of Cosmology and Gravitation, University of Portsmouth, Portsmouth, PO1 3FX, UK}

\author[0000-0002-8262-2924]{Michael W. Coughlin}
\affiliation{School of Physics and Astronomy, University of Minnesota, Minneapolis, Minnesota 55455, USA}

\author[0000-0001-6586-4297]{Ekaterine Dadiani}
\affiliation{\mcwilliams}

\author[0009-0004-2836-1059]{Veronica Diaz}
\affiliation{\mcwilliams}

\author[0000-0002-3168-0139]{Matthew J. Graham}
\affiliation{Division of Physics, Mathematics, and Astronomy, California Institute of Technology, Pasadena, CA 91125, USA}

\author[0000-0003-3270-7644]{Daniel Gruen}
\affiliation{\usm}
\affiliation{Excellence Cluster ORIGINS, Boltzmannstr. 2, 85748 Garching, Germany}

\author[0009-0000-4830-1484]{Keerthi Kunnumkai}
\affiliation{\mcwilliams}

\author[0000-0003-0738-8186]{Jake Postiglione}
\affiliation{Department of Science, BMCC, City University of New York, New York, NY 10007, USA}

\author[0000-0002-5466-3892]{Arno Riffeser}
\affiliation{\usm}
\affiliation{Max Planck Institute for Extraterrestrial Physics, Giessenbachstr. 1, 85748 Garching, Germany}

\author[0000-0002-1154-8317]{Julian S. Sommer}
\affiliation{\usm}

\author[0000-0001-5567-1301]{Francisco Valdes}
\affiliation{NSF National Optical-Infrared Research Laboratory, 950 N. Cherry Ave., Tucson, AZ 85719, USA}


\begin{abstract}
We carried out long-term monitoring of the LIGO/Virgo/KAGRA binary black hole (BBH) merger candidate S230922g in search of electromagnetic emission from the interaction of the merger remnant with an embedding active galactic nuclei (AGN) accretion disk.
Using a dataset primarily composed of wide-field imaging from the Dark Energy Camera (DECam) and supplemented by additional photometric and spectroscopic resources, we searched $\sim 70\%$ of the sky area probability for transient phenomena, and discovered 6 counterpart candidates.
One especially promising candidate - AT 2023aagj - exhibited temporally varying asymmetric components in spectral broad line regions, a feature potentially indicative of an off-center event such as a BBH merger.
This represents the first \emph{live} search and multiwavelength, photometric, and spectroscopic monitoring of a GW BBH optical counterpart candidate in the disk of an AGN.
\end{abstract}

\section{Introduction}
\label{sec:intro}

As multimessenger astronomy continues to develop, the branch of this field dedicated to studying \glspl{cbc} stands as an important forerunner as the first to include direct observations of \glspl{gw} \citep{abbottMultimessengerObservationsBinary2017}.
These events manifest phenomena at scales presently unobtainable in man-made environments, and hence are important natural laboratories in which to study extreme physics such as heavy element nucleosynthesis \citep{symbalistyNeutronStarCollisions1982} and measure key physical quantities such as the expansion rate of the universe \citep{schutzDeterminingHubbleConstant1986, holzUsingGravitationalWaveStandard2005}.

This potential was realized following the first successful multimessenger detection of the binary neutron star merger GW170817/GRB 170817A \citep{abbottGravitationalWavesGammaRays2017}, wherein a global electromagnetic astronomy campaign succeeded in locating a \gls{kn} \citep{coulterSwopeSupernovaSurvey2017,Arcavi2017,Andreoni2017,Drout2017,Nicholl2017,Evans2017,Lipunov2017,Pian2017,Smartt2017,Utsumi2017,Valenti2017,kasliwal2017illuminating,troja2017x,Margutti2017,Shappee2017,soares-santosElectromagneticCounterpartBinary2017,Chornock2017,Tanvir2017} incident with the initially identified \gls{gw} and \gls{grb} events \citep{abbottMultimessengerObservationsBinary2017,Savchenko2017,Goldstein2017}. 
This single event by itself was enough to result in hundreds of analyses, including constraints on \gls{ns} physics \citep[e.g.][]{abbottGW170817MeasurementsNeutron2018} and the Hubble constant $H_{\rm 0}$ \citep{abbottGravitationalwaveStandardSiren2017}. 
In the case of $H_{\rm 0}$, it is predicted that $\mathcal{O}(100)$ \gls{bns} events with \gls{em} counterparts are required to make a $\sim$few\% measurement of the parameter \citep{chenTwoCentHubble2018}, although joint multimessenger constraints on the binary viewing angle can further improve precision by a factor of a few (e.g. \citealt{Hotokezaka2019, palmese2024standard}).

\begin{figure*}
    \centering
    \gridline{
        \fig{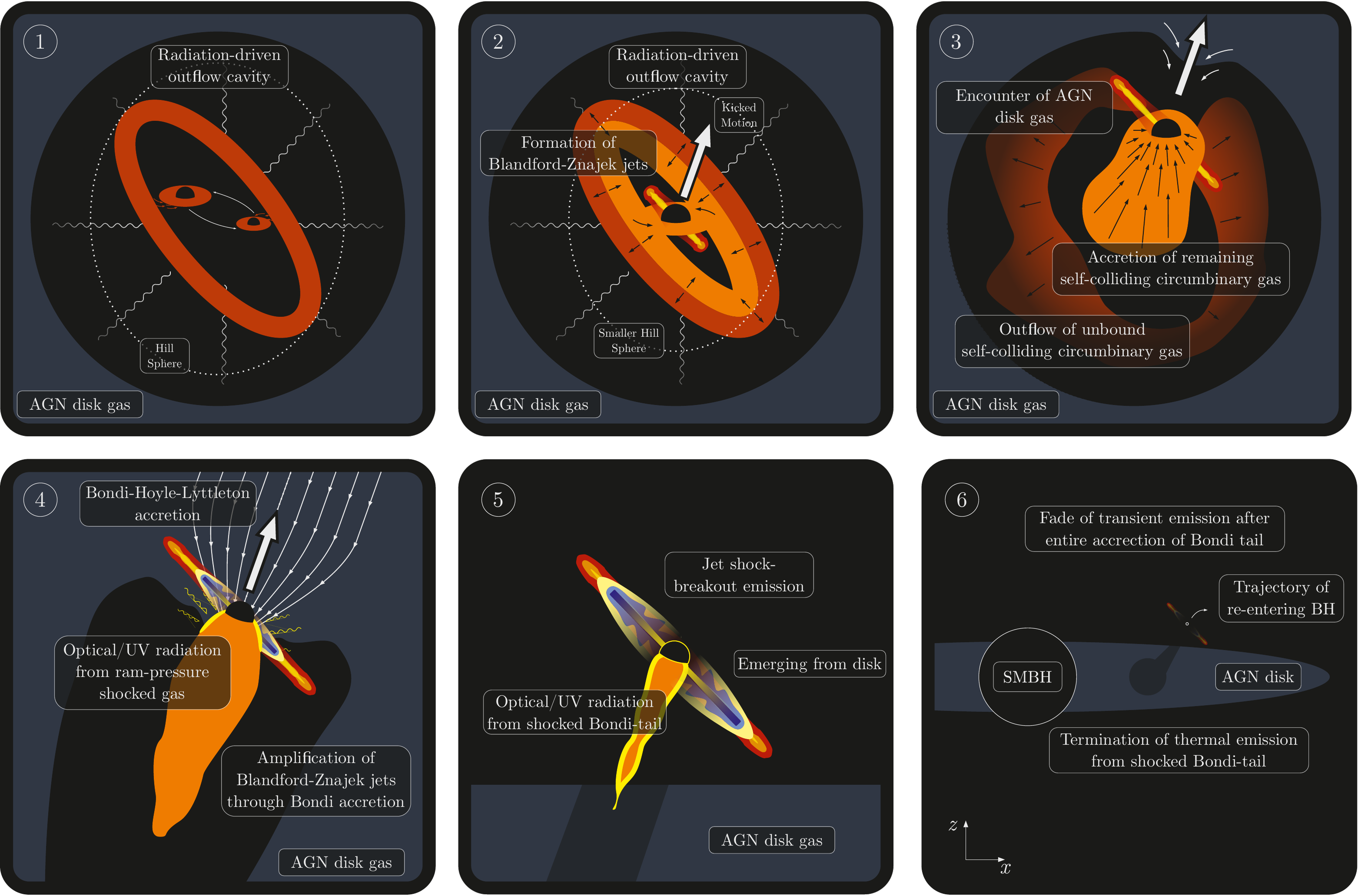}{0.95\textwidth}{}
    }
    \caption{
         Multi-panel schematic showing the mechanism believed to underpin luminous \gls{em} counterparts to \gls{bbh} mergers in \gls{agn} disks.
         In panel 1, the pre-merger \gls{bbh} accretes from mini-disks within its Hill sphere in the \gls{agn} disk midplane and blows a cocoon within the disk via feedback.
        In panel 2, the merger happens, forming a highly spinning \gls{bh} (dimensionless spin parameter $a \sim 0.7$ typically).
        A jet is presumed to form at this stage (although it has yet to be established whether such a jet can persist for long, or whether it is choked off by high mass accretion).
        Mass and spin asymmetries in the progenitor black holes lead to a kick at merger (depicted by the arrow in panel 2).
        Panels 3 and 4 show the development of \gls{bhl} accretion as the newly merged \gls{bh} exits its original Hill sphere into the rest of the \gls{agn} disk, powering a luminous transient.
        In panel 5 the \gls{bh} emerges from the \gls{agn} disk, dragging disk gas with it.
        In panel 6 the \gls{em} emission fades as the disk material is consumed and the \gls{bh} continues on an inclined orbit around the \gls{smbh}, and will re-enter the \gls{agn} disk on half the orbital timescale.
    }
    \label{fig:model}
\end{figure*}

While \gls{bns} mergers are the only class of \gls{gw} event with a confirmed \gls{em} counterpart, it is also predicted that \gls{bbh} mergers can produce counterparts in certain circumstances.
The majority of proposed counterpart mechanisms involve the interaction of the binary or merger remnant with a coexistent medium, usually the gaseous accretion disk of an \gls{agn}, whether through accretion \citep{bartosRapidBrightStellarmass2017}, ram-pressure stripping of gas about the remnant or jetted Bondi accretion \citep{mckernanRampressureStrippingKicked2019}, \response{or breakout emission from accretion \citep{2021ApJ...916..111K, 2024arXiv240709945R}} or a post-merger jet \citep{tagawaObservableSignatureMerging2023,Tagawa:2023gpi}. 
A cartoon of the \gls{agn}-associated \gls{em} counterpart mechanism is show in Figure~\ref{fig:model}.

Counterpart candidates for \gls{bbh} mergers from previous \gls{lvk} observing runs have been proposed \citep{grahamCandidateElectromagneticCounterpart2020, grahamLightDarkSearching2023}, although confirmation of a counterpart has remained challenging.
There are several reasons for this difficulty.
First, even if we know that a given merger happens in an \gls{agn}, in principle there is only a $\mathcal{O}(1/4)$ chance that we could detect an \gls{em} counterpart: our view of around half of \gls{agn} (the Type 2s) is obscured, and an unfortunate kick-direction out the opposite side of a Type 1 \gls{agn} could still obscure any emission \citep{grahamLightDarkSearching2023}.
Second, even if a flare emerges from an \gls{agn}, it must be discerned in the presence of other \gls{agn} variability.
This makes our task of searching for counterparts more challenging in brighter \gls{agn}, and our search is biased against less luminous counterparts.
Third, possible flare parameters are weakly constrained, both because \gls{agn} disk properties are uncertain to orders of magnitude, and because the properties of the post-merger flare depend on these uncertain parameters.

One possible means to confirm a counterpart is through optical spectroscopy: because the \gls{bbh} merger occurs off-center in the \gls{agn} disk, the resulting flare could unevenly illuminate the \gls{blr} of the \gls{agn}, resulting in asymmetric broad-line features in the optical spectrum \citep{mckernanRampressureStrippingKicked2019}.
Synchronized evolution of a transient with such spectral features can be a key piece of evidence in favor of the classification of a transient as a \gls{em} counterpart to a \gls{bbh} merger.
At the very least, an evolving asymmetric line profile suggests an off-center flaring event in the \gls{agn}, ruling out most sources of \gls{agn} variability near the \gls{smbh}.
Flare energetics and lightcurve profiles can allow us to rule out e.g. embedded supernovae, leaving few candidates for a sufficiently energetic off-center \gls{agn} flare, including a \gls{bbh} merger.

Certainly, the search for \gls{em} counterparts to \gls{bbh} mergers has unique challenges versus the search for \gls{ns} merger counterparts (intrinsic \gls{agn} variability, delay time uncertainties, etc.); however, as the historical rate of detection of \gls{bbh} mergers has been $\sim\mathcal{O}(100)\times$ greater than that for \gls{ns} mergers, there are many more opportunities to search for a counterpart to the former kind of event.
Because multimessenger observations of \gls{bbh} mergers are also useful in making cosmological measurements \citep{palmeseLIGOVirgoBlack2021,bomStandardSirenCosmology2023,Alves_2024}, the pursuit of \gls{em} counterparts to \gls{bbh} mergers has the potential to significantly contribute to the multimessenger observations required to make the first 2\% measurement of the Hubble constant with standard sirens, which the level of precision needed to help us understand the Hubble tension \citep{chenTwoCentHubble2018}.

Currently, \gls{gw} follow-up of distant events is best enabled with technology capable of meeting the colloquial requirements of ``wide, fast, deep", as to ensure the rapid and thorough coverage of event volumes.
The \gls{decam} \citep{flaugherDarkEnergyCamera2015} on the 4m Victor M. Blanco Telescope st the \gls{ctio} is one of the premiere instruments that can address these requirements.
\gls{decam} already has a respectable history in \gls{gw} follow-up, having been used for this purpose from the first \gls{gw} event \citep{soares-santosDARKENERGYCAMERA2016, annisDARKENERGYCAMERA2016} and many events since then \citep{goldsteinGROWTHS190426cRealtime2019a,andreoniGROWTHS190510gDECam2019a,hernerOpticalFollowupGravitational2020,andreoniGROWTHS190814bvDeep2020,morganConstraintsPhysicalProperties2020,garciaDESGWSearchElectromagnetic2020}, and importantly being one of the first instruments to detect the counterpart \gls{kn} to GW170817 \citep{soares-santosElectromagneticCounterpartBinary2017}.
The new survey program \gls{gwmmads} (PIs: Andreoni \& Palmese) is designed to find \gls{em} counterparts to \gls{bns}, \gls{ns}\gls{bh}, and \gls{bbh} mergers via rapid DECam follow-up of \gls{gw} events during \gls{o4}.

In this work, we present results from our follow up of the \gls{gw} event \gweventid.
\gweventid is a \gls{lvk} \gls{bbh} merger candidate detected at 2023-09-22 02:03:44.886 UTC \citep{ligoscientificcollaborationLIGOVirgoKAGRA2023, ligoscientificcollaborationLIGOVirgoKAGRA2023a}.
The 90\% credible region of 324 deg$^2$ is one of the smallest areas of \gls{o4}a (the first part of O4, completed in January 2024); this, combined with the localization being visible from Chile in the months following the trigger, made an effective follow-up with \gls{decam} possible in pursuit of potential \gls{agn}-linked counterparts like those mentioned previously. 
In this paper we present our long-term monitoring of \gweventid; the paper is organized as follows:
\S\ref{sec:data} presents our program's generic strategy for \gls{gw} follow-up and the data collected, highlighting components especially relevant for \gweventid; \S\ref{sec:vetting} describes the methods we used to distill our population of detected transients into a shortlist of counterpart candidates; \S\ref{sec:candidates} describes our most favored candidate, and to a lesser extent additional candidates of interest; in \S\ref{sec:discussion} we estimate parameters for our most favored counterpart candidate, finding it well within the confines of existing \gls{bbh} counterpart theory; in \S\ref{sec:conclusions} we summarize our findings and highlight considerations relevant for future efforts.
Throughout this work we use a flat $\Lambda$CDM model with $H_0 = 70$ km/s/Mpc and matter density of $\Omega_m = 0.3$,

\section{Data}\label{sec:data}

\subsection{GW data}

S230922g is an event of interest because it was detected with high significance (False Alarm Rate - FAR - 1 per $1.6\times10^{-16}$ years from \texttt{gstlal}; \citealt{Messick_2017gstlal,sachdev2019gstlal}) by both LIGO Livingston and LIGO Hanford, it is well localized compared to the O4a population, and it has $\sim$100\% probability of being a \gls{bbh}.
It was also identified with high significance by the Burst \texttt{CWB} \citep{Klimenko_2016} search pipeline, potentially indicating a loud, short burst, as one may expect for a massive BBH spending a short fraction of the late inspiral phase in the LVK band.
The luminosity distance of the event, marginalized over the entire sky, is $d_L=1491 \pm 443$ Mpc.
These quantities are reported with the skymap from the \texttt{Bilby} \citep{Ashton_2019} reduction \citep{2023GCN.34758....1L}; we use this skymap for our work.

Higher mass ($M_{\rm tot} \gtrsim 50~M_\odot$) \gls{bbh} mergers are more likely to have originated from the disks of \glspl{agn} (e.g. \citealt{Gayathri_2021}), and so we also inform our decision to trigger \gls{bbh} follow up by estimating the total mass of the binary.
We follow a similar calculation to that in \citet{grahamCandidateElectromagneticCounterpart2020}, and consider that $A_{90}\propto~{\rm SNR}^{-2}$ \citep{Berry_2015}, as well as ${\rm SNR}\propto \mathcal{M}_c^{5/6}/d_L$ \citep{Finn_1993}, where $\mathcal{M}_c$ is the chirp mass of the binary.
For this event, we assume a 1.4-1.4 $M_\odot$ binary neutron star merger detection horizon of 150 and 152 Mpc for Hanford and Livingston, respectively, an equal mass system, and the luminosity distance marginalized over the sky; we derive a total rest frame mass of $M_{\rm tot} \sim 90~M_\odot$ for \gweventid.
\response{The BNS merger detection horizon serves as an additional scaling factor in the overall calculation.} Note that this estimate is highly uncertain with an error bar of at least a factor of 2. \response{This uncertainty is informed from the calibration of the proposed mass relationship with the LIGO O3 BBH data set.}

\subsection{DECam observations}\label{subsec:observations}

Our team was notified of \gweventid through a \gls{gcn} listener\footnote{Adapted from \url{https://github.com/scimma/slackbot}.} that sent a digest of the event to our Slack workspace.\footnote{Our team also uses the Fritz science data platform \citep{coughlinDataSciencePlatform2023} for notifications of \gls{gw} events via phone call, but we reserve this kind of notification for exceptionally time-sensitive events such as \gls{bns} and \gls{ns}\gls{bh} mergers.}
Our first response was to generate an initial observing strategy with \gwemopt \citep{coughlinOptimizingSearchesElectromagnetic2018} to asses the probability coverage possible with a single night of \gls{decam} observations.
\gwemopt selects exposure pointings from a preset sky tiling.
Because our image subtraction pipeline uses archival \gls{decam} exposures as templates, we implemented a custom tiling based on images calibrated by the \gls{noirlab} \gls{cp} \citep{2014ASPC..485..379V} and made available through the NOIRLab Astro Data Archive\footnote{\url{https://astroarchive.noirlab.edu/}} \citep{2021Mirro...2...33M} to ensure our observing plans include pointings with usable templates.
During each observing night, the \gls{cp} calibrates the response and astrometry of each exposure shortly after it is taken which is then available to the image subtraction pipeline.

\begin{figure}
    \centering
    \includegraphics[width=0.49\textwidth]{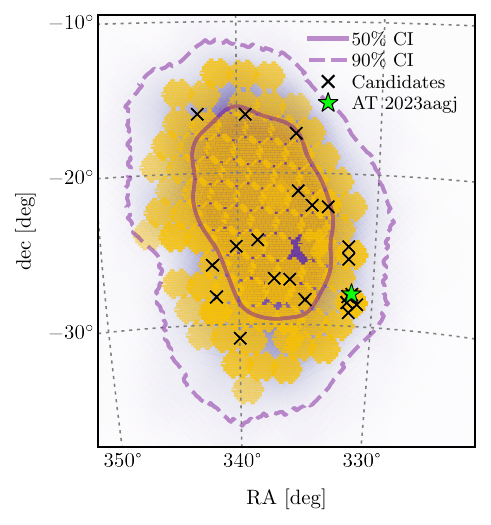}
    \caption{
        Our observation plan for follow up of \gweventid.
        The \gls{lvk} skymap for the event is plotted on the lower layers of the figure, and the 50\% (90\%) credible interval regions are outlined in the solid (dashed) line.
        The candidates composing our shortlist are shown as black X's, with our favored candidate indicated with a green star.
    }
    \label{fig:plot_followup}
\end{figure}

We determined that 70\% of the \gls{gw} localization area could be covered with 60\,s and 80\,s \gls{decam} exposures in $g$- and $i$-bands, respectively, with $\sim$4.7 hours of telescope time.
Our \gls{decam} tiling is shown in Figure~\ref{fig:plot_followup}, overlaid on the \gls{lvk} Bibly skymap \citep{ligoscientificcollaborationLIGOVirgoKAGRA2023a} for \gweventid.
When this plan was finalized, we communicated our pointings to the larger astronomical community though the \gls{gw} Treasure Map\footnote{\url{https://treasuremap.space/}} \citep{Wyatt_2020}, and we updated information appropriately as we executed our plan.

\response{Because most \gls{bbh} counterpart models concern the emergence of an \gls{em} signature from the interior of an \gls{agn} disc, delays of $\mathcal{O}(10-100)$ days are expected before such a signature becomes observable}.
Even so, it is valuable to initiate \gls{em} follow up for these kinds of events within a few days of the \gls{gw} trigger, as high-cadence archival data are generally not available in the localization area of a given \gls{gw} event and a baseline of sources in the localization area can be useful when vetting candidates.
Accordingly, we decided to trigger our \gls{too} program for the night following the event (the evening of September 22, 2023) in order to establish a baseline of activity at the \gls{gw} merger time and to facilitate the identification of novel phenomena in subsequent observing epochs.

On the first night we were only able to complete $\sim$1 hour of observations before inclement weather began, but we were able to observe our full plan in its entirety the following night (September 23).
We publicly reported transients in the area through  \gls{gcn} within 24 hours of these observations \citep{2023GCN.34763....1C}.
Further observations were conducted 11, 33, 41, and 70 days after the \gls{lvk} trigger (October 3, October 25, November 2, and December 1).
As we processed the data throughout the campaign, we took additional observations of the most interesting candidates, adding up to eight epochs altogether.
We submitted all transient sources discovered through this campaign to the Fritz SkyPortal \citep{skyportal2019,coughlinDataSciencePlatform2023}.
After composing our final candidate shortlist (as detailed in \S\ref{sec:vetting}), we reported the respective sources to the \gls{tns}.

\subsection{Spectroscopic observations}

\gls{bbh} counterpart models have some degeneracy with those for stochastic \gls{agn} variability and other transient events like \glspl{tde}, but spectroscopic data has been predicted to serve as a key discriminator among these phenomena: specifically, the off-center location of a \gls{bbh} merger in the accretion disk of the \gls{agn} as opposed to the central location of \glspl{tde} or accretion-based phenomena is expected to induce asymmetry in the broad lines of \gls{agn} spectra as the transient event ``washes" over the \gls{blr} in an asymmetric manner \citep{mckernanRampressureStrippingKicked2019}.
The detection of a \gls{blr} asymmetry that evolves in concert with the light curve of a transient is a smoking gun that locates the transient in an off-center position in the accretion disk expected for \gls{bbh} mergers, and the coincidence of such an event with a \gls{gw} merger is a strong evidence standard astronomical techniques can provide linking \gls{bbh} mergers and \gls{agn} flares.

In pursuit of such evidence, we triggered several spectroscopic resources for additional follow-up of transients whose light curves demonstrated proposed counterpart features (see \S\ref{subsec:autovet}).
A total of five spectra were collected during our follow-up campaign: two with Gemini GMOS, and one each with Keck LRIS, SALT RSS, and P200 DBSP.
We discuss the methodology for each spectrum in the context of the respective candidate in \S\ref{subsec:othercands}.


\section{Methods}
\label{sec:vetting}

\subsection{Automated vetting}
\label{subsec:autovet}

\begin{deluxetable}{ccc}
    \tablecaption{
        Summary of selection cuts applied to our pipeline products.
        The table is split into two sections: cuts applied to individual difference image detections (first 3 cuts), and those applied to multi-epoch light curves composed of coincident detections (remaining 6 cuts).
        \label{tab:vetting}
    }
    \tablehead{\colhead{Vetting filter} & \colhead{\# passed} & \colhead{Frac. passed}}
    \startdata
        \multicolumn{3}{c}{\textit{Detection-based cuts}} \\
        \hline
        Initial photometry detections & 25,392,140 & 1.00 \\
        Data quality masking & 14,053,259 & 0.553 \\
        Real/bogus score $\ge$ 0.7 & 1,501,800 & 0.059 \\
        \hline 
        \multicolumn{3}{c}{\textit{Source-based cuts}} \\
        \hline
        Initial transients & 233,313 & 1.00 \\
        Remove LS variable stars & 140,230 & 0.601 \\
        Remove MPC objects & 18,528 & 0.079 \\
        $\ge$1 real/bogus score $\ge$0.9 & 17,515 & 0.075 \\
        $\ge$2 detections with $\Delta t \ge 30$ min & 3558 & 0.015 \\
        LS galaxy sep $\leq$ 1" & 2388 & 0.010 \\
        Additional vetting & 6 & 2.6e-5
    \enddata
\end{deluxetable}

We analyze our data with our difference photometry pipeline described in \citet{pipeline}.
We perform image subtraction with \gls{sfft} \citep{huImageSubtractionFourier2022}, a scalable image subtraction algorithm and GPU-enabled implementation of the same that produces difference images up to an order of magnitude faster than widely used tools with comparable accuracy.
Aperture photometry is conducted on the resulting difference images using \texttt{SExtractor} \citep{bertinSExtractorSoftwareSource1996a} with fluxes calibrated to the \gls{ls} source catalog and corrected for extinction using the \texttt{dustmaps} package \citep{greenDustmapsPythonInterface2018} and the $E(B-V)_{\rm SFD}$ coefficients from \citealt{schlaflyMEASURINGREDDENINGSLOAN2011}.
Our image differencing pipeline detected over 25 million transient features in our campaign data, which we distilled into a tractable list of astrophysical transients through a series of cuts.
Table \ref{tab:vetting} summarizes the vetting steps we applied, which are described in this section.

We first remove any detections on difference images contaminated by bad pixels recorded in the \gls{decam} \gls{cp} data quality mask products.
Surviving features were then scored using a \response{rotation-invariant \gls{cnn} \citep{2017ApJ...836...97C, 2022FrASS...9.7100S}} trained on archival \gls{decam} data products from our pipeline to perform real/bogus classification.
Possible \gls{cnn} scores range from 0 to 1, with higher scores associated with more realistic astrophysical objects.
\response{From testing on archival \gls{decam} data a threshold of 0.7 was found to facilitate a 98\% recall rate with a 7.2\% bogus contamination rate; this threshold was accepted as our cutoff for this analysis.}
Roughly 6\% of the initial list of detections passed these steps to form our list of real astrophysical detections.

These detections were then cross-matched among themselves to identify features present in data from multiple epochs and filters; the resulting list of transients consisted of 233,313 objects.
This list was subsequently crossmatched with the \gls{ls} DR10.1 catalog \citep{Dey2019} and the \gls{mpc} service, and any transients respectively matched to star-like sources (defined as a source assigned a \gls{ls} morphological type of ``PSF") or minor planets were removed from our search.
The \gls{ls} crossmatch was also used to sort the sources into three categories based on proximity to the nearest galaxy-type\footnote{We consider all \gls{ls} sources with PSF types other than ``PSF" (stellar) and ``DUP" (extended source components) as ``galaxy-type" sources.} LS source: sources within 0.3” of a galaxy-type source were labeled as ``A" sources,
sources greater than 1.0” away from the nearest such source were labeled as ``T" sources, sources falling in the middle ground between these two categories were labeled as ``C" sources.

Two final cuts were then applied to limit our search to the most realistic persistent sources, requiring sources to have least two distinct observations separated by $>$30 minutes, and at least one high-fidelity detection (real/bogus \gls{cnn} score $\ge$0.9).
Of the remaining 3558 sources, 2388 of them were within 1" of a galaxy-type \gls{ls} object (``A" and ``C" sources); this final group composed our automated list of candidates based on our photometric \gls{decam} data.
See Table \ref{tab:vetting} for a summary of cuts applied and resulting numbers of candidates at each cut.

\subsection{Additional vetting} 
\label{subsec:handsvet}

Our 2388 automatically-identified sources were then examined by eye to identify candidates for further study.
Sources flagged for further investigation included those that had brightened since the time of the \gls{gw} event and did not exhibit early reddening (this latter criteria was motivated by the expectation that the luminous counterpart emerges from the depths of the accretion disk, see Figure~\ref{fig:model}, and we do not expect it to redden as the optical depth decreases).
Other considerations included whether there was a perceptible delay time from the \gls{gw} event time (reflective of an \gls{em} counterpart needing time to escape the embedding accretion disk before being observed) and whether sources appeared similar to different known transients such as \glspl{sn}.
Image stamps were also examined for each source to exclude any artifacts and the like that survived the \gls{cnn} cut.

The resulting list was trimmed to remove any transients over 2$\sigma$ away from the \gls{gw} distance posterior, as such information was available.
To determine distances to each transient, the \gls{ls} galaxy-type object matched to each transient was cross-matched with several galaxy catalogs (with a search radius of 1") in search of a redshift measurement.
Galaxy catalogs used included the \gls{ned} Local Volume Sample \citep{cook2023completeness} \footnote{\url{https://ned.ipac.caltech.edu/}}, the \gls{desi} galaxy catalog \citep{desiedr}, and Quaia \citep{storey-fisherQuaiaGaiaunWISEQuasar2024}; \gls{ls} DR10.1 photometric redshifts were also included.
If the host galaxy matched in multiple catalogs, then the best redshift measurement was used, preferring spectroscopic redshifts over Quaia ``spectrophotometric" redshifts \citep{storey-fisherQuaiaGaiaunWISEQuasar2024}, and the latter over photometric redshifts.
Comparing the available redshift measurements to the distance posteriors (taken from the skymap HEALPix\footnote{\url{https://healpix.sourceforge.io/}} tile containing the respective transient) informed the elimination of sources beyond the 2$\sigma$ threshold cut, using the larger of the two uncertainties between the \gls{gw} and galaxy catalog measurements.
Note that transients whose host galaxy lacked any redshift measurement were not subject to this cut.

Candidates in our shortlist were photometrically classified with \gls{parsnip} \citep{booneParSNIPGenerativeModels2021}, assuming a model trained on the PLAsTiCC \citep{Kessler_2019} simulations; because \gls{parsnip} uses redshift as an input parameter, this classification was limited to only those sources with a redshift measurement.
When classifying a light curve, \gls{parsnip} assigns a probability to each of a set list of transient classes so that  the total probability sums to 1, with each probability reflecting the relative likeness of the light curve to each class.
We note that \gls{parsnip} does not contain a catch-all class such as ``Other" for use when classifying transients of unfamiliar phenomenology: that is, if a transient is strongly dissimilar to all but one of the classes, it could receive a high classification probability for that one class, even if it not a strong match in and of itself.
We consider the ``TDE" \gls{parsnip} class as the one most similar to the \gls{agn} flares of interest (all other classes concern some kind of \gls{sn}, except for the \gls{kn} class, whose typical timescales are considerably shorter than those we are interested in).
Accordingly, in our use case we broadly interpret transients classified as ``TDE" as those that are not identifiable as \gls{sn}-like events, and we refer to the ``TDE" class as ``Non-SN" to better reflect this perspective.
We find that most of our classified transients receive this classification.

The final shortlist of 23 transients surviving these cuts is shown in Table \ref{tab:candidates}.
We separate several subsets from this list to identify transients that we exclude from our search via asynchronous analysis after our follow-up campaign; these subsets appear in labeled sections of the table.
6 of the transients in our shortlist showed significant brightening, but did not peak during the time they were observed (so potentially consistent with longer-term AGN variability), and so we are unable to consider their full nature with our present dataset.
9 transients, while initially interesting for further monitoring, exhibited reddening in later epochs, and so are excluded through disagreement with our assumed counterpart model that predicts a signature that becomes more blue with time.
Finally, 2 transients were excluded as counterpart candidates through spectroscopic classification (see \S\ref{subsubsec:AT2023aden} and \S\ref{subsubsec:AT2023uab}).
The remaining 6 transients are those that cannot be excluded as counterparts to \gweventid, and are listed at the top of the table.
Table \ref{tab:candidates} includes host redshift information (where available), along with \gls{gw} skymap localization information: the 1$\sigma$ distance posterior expressed as redshift under the assumed cosmology, and the 2D and 3D \gls{ci} as calculated with the \texttt{ligo.skymap.postprocess.crossmatch.crossmatch} routine\footnote{\url{https://lscsoft.docs.ligo.org/ligo.skymap/postprocess/crossmatch.html\#ligo.skymap.postprocess.crossmatch.crossmatch}} \citep{singerGoingDistanceMapping2016}.
The transients are ranked by ascending 2D \gls{ci}, s.t. the objects in the highest probability regions are listed first.
The highest-probability \gls{parsnip} class and the associated probability are listed in the last two columns of the table, where available.

\input{candidates_table}

\section{Candidate counterparts}
\label{sec:candidates}

In this section we describe candidate counterparts that received particular scrutiny during our campaign, including one we identify as the most likely to be a counterpart to \gweventid.
While we generally refer to candidates by their \gls{tns} name, we include the internal name for each of our transients as well, as these names were used for initial reports.
Internal names are assembled from the date of discovery and RA/dec of the source; for example, the internal name of C202309242206400m275139 refers to a transient first detected on September 24, 2023 with an RA/dec of 22h 06m 40.0s, -27$^\circ$ 51m 39s (the letter between the RA and dec can be either ``p" or ``m", denoting a positive or negative declination, respectively).
The leading character in the object names indicates the proximity of the transient to the nearest galaxy-type \gls{ls} source, in the same convention explained in \S\ref{subsec:autovet} (``A" for nuclear sources, ``T" for non-nuclear sources, and ``C" for marginally nuclear sources).
The internal names for all transients in our shortlist can be found in the subfigure titles of Figure~\ref{fig:light_curves_other}.

\subsection{AT 2023aagj (C202309242206400m275139)}
\label{subsec:AT2023aagj}

\begin{figure*}
    \centering
    \includegraphics[width=0.95\textwidth]{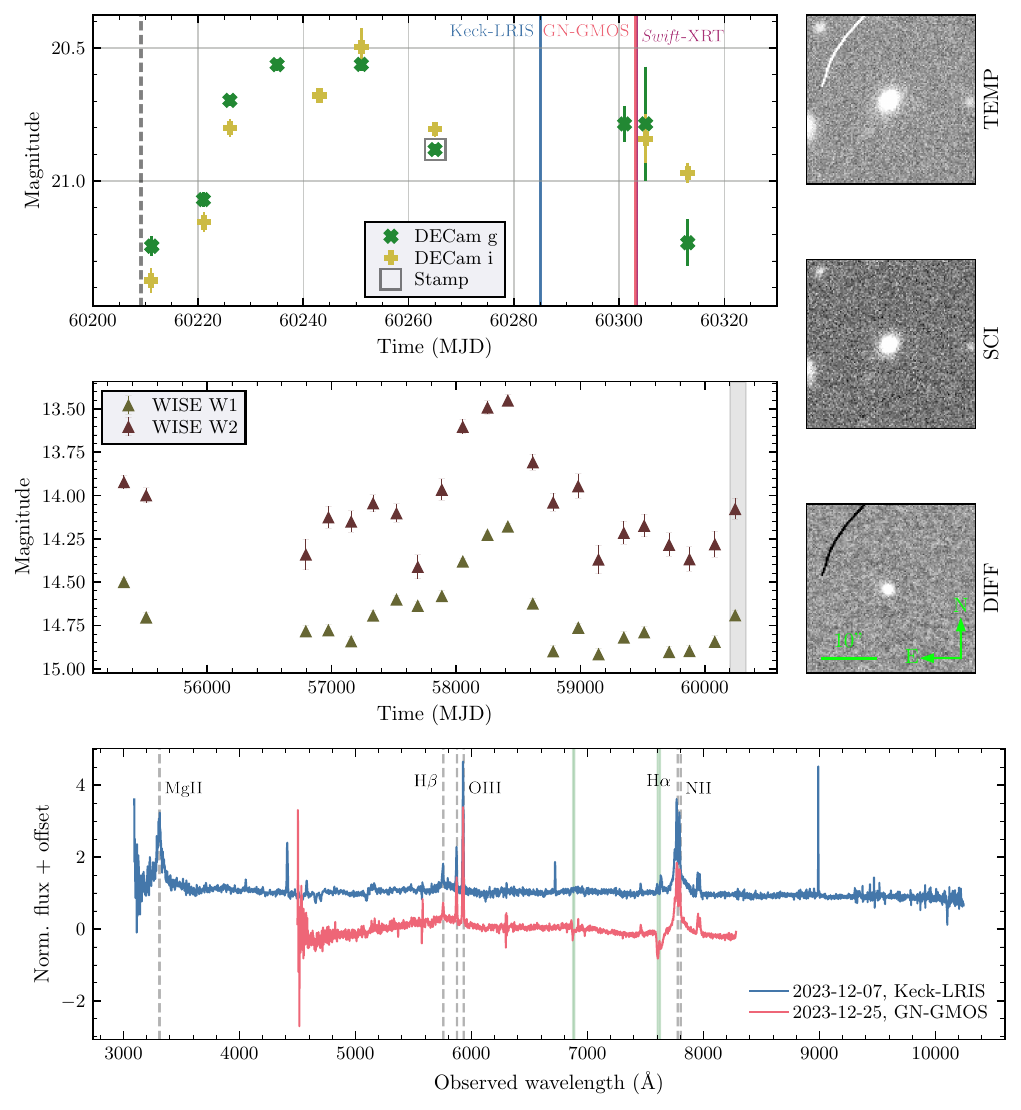}
    \caption{
        Observational data for AT 2023aagj.
        \textit{Upper left:} Difference photometry for the transient.
        $g(i)$-band data are shown in green (yellow). 
        \response{The dashed gray line marks the GW event time of S230922g, and} the colored vertical lines indicate the times additional data (spectra and X-ray) were taken.
        \textit{Upper right:} Stamps from the template, science, and difference images for a sample epoch; the epoch from which the sample stamps are taken is outlined in the photometry plot with a gray square.
        \textit{Center left:}
        WISE photometry for the host of AT 2023aagj.
        The span of the difference photometry plot is shaded in gray.
        \textit{Bottom:} Spectra taken for the transient.
        Telluric regions are shaded in green.
        Several known emission lines (shifted according to the measured redshift of the host) are labeled and shown as dashed gray lines.
    }
    \label{fig:C202309242206400m275139}
\end{figure*}

We consider AT 2023aagj to be the transient from our sample most likely to be an optical counterpart to \gweventid.
This transient is distinguished by a $\sim$1 magnitude brightening over a period of a month, remaining blue in color throughout its evolution.
It is within 0.5" of the host galaxy centroid, and as such may be associated with activity in the host nucleus.
Difference photometry, spectra, and sample stamps for this candidate are visible in Figure~\ref{fig:C202309242206400m275139}.
\gls{parsnip} classifies AT 2023aagj as a ``TDE"/``Non-SN"-type object with a probability of 94.4\%, which in this context distinguishes the transient from \glspl{sn} and the like.

\begin{figure*}
     \centering
     \gridline{
        \includegraphics[width=0.49\textwidth]{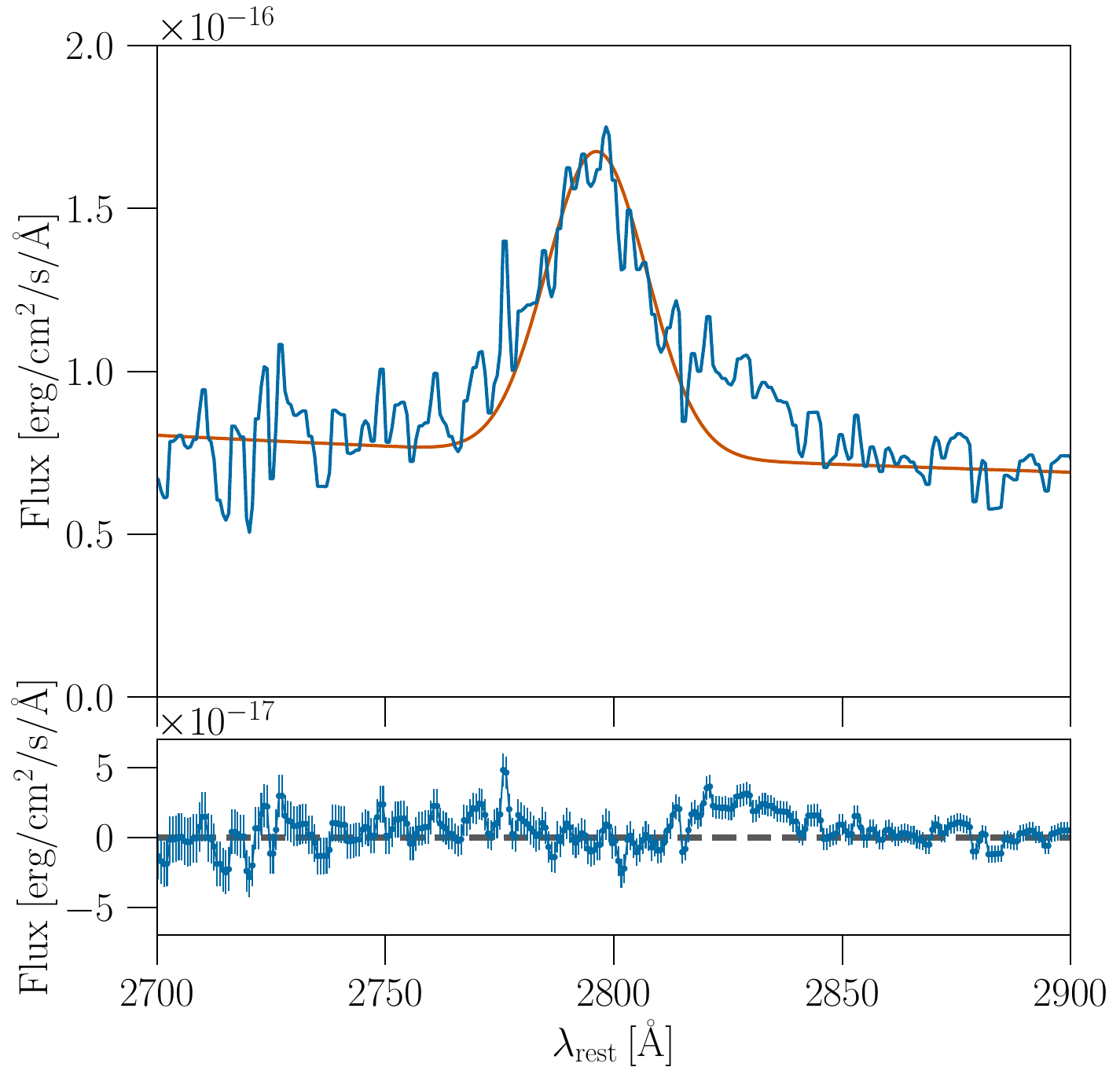}
        \includegraphics[width=0.49\textwidth]{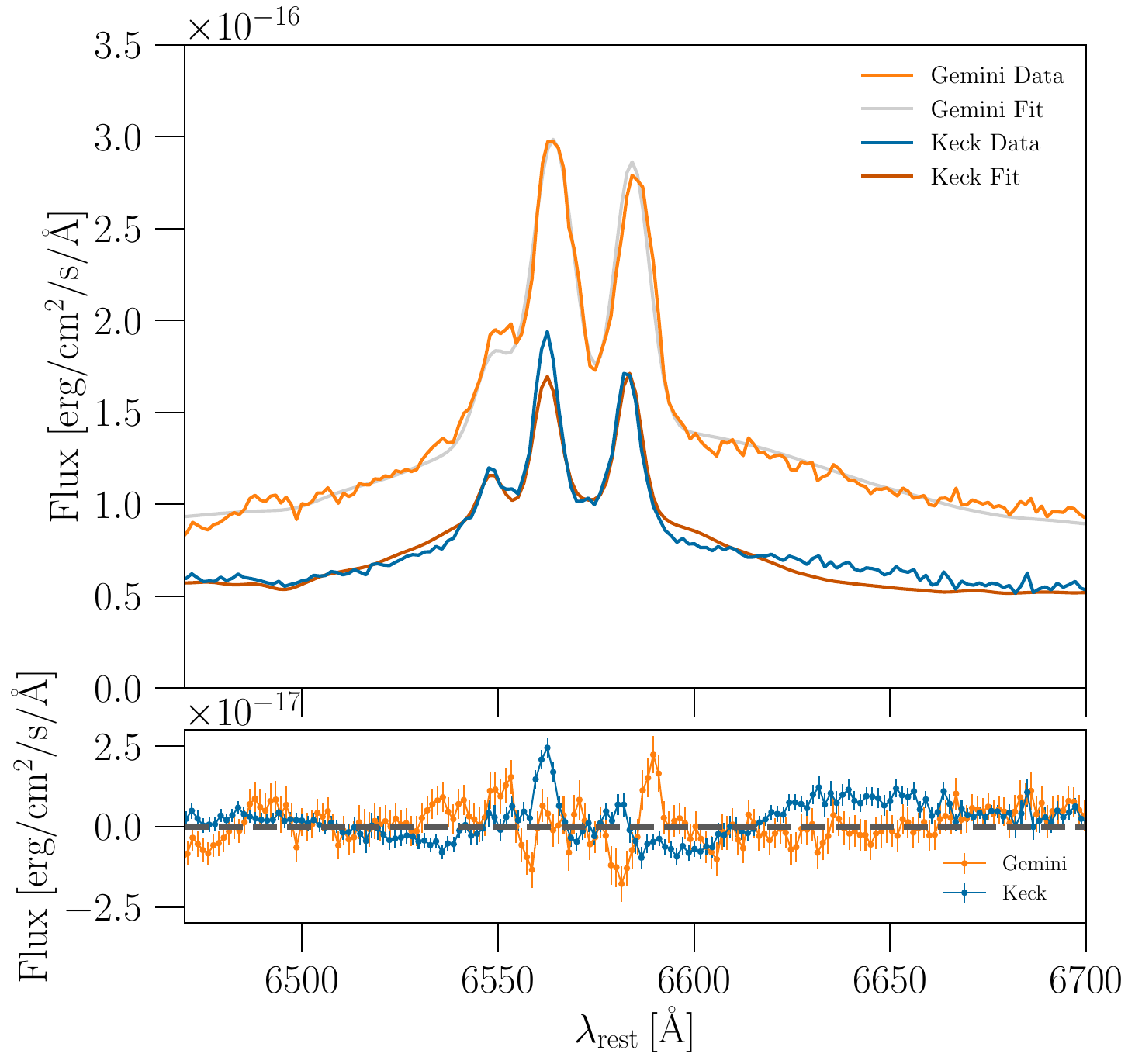}
     }
     \caption{
        \texttt{pPXF} model and data for the two AT 2023aagj spectra around the two broad lines of interest.
        The bottom panels show the residuals between the model and data. An asymmetry redward of the line appears to be present in the Keck spectrum, taken closer to peak, but disappears in the later Gemini spectrum. }
    \label{fig:asymmetry}
\end{figure*}

Two spectra of AT 2023aagj were taken before the transient faded.
The initial Keck \gls{lris} spectrum, taken on December 7 (PI: Kasliwal, PID: C360), and reduced with the standard LPIPE routine \citep{lpipe}, revealed ionized gas emission lines and broad line features for the H$\alpha$ and MgII regions, indicating a possible \gls{agn} host.
The redshift of the host was calculated to be 0.184, placing the transient within the 83\% credible volume for the \gweventid skymap and consistent at the 1.4$\sigma$ level with the luminosity distance posterior of the \gls{gw} event.
An additional spectrum of the object was taken with Gemini North \gls{gmos} on December 25, 2023 (PI: Cabrera, PID: GN-2023B-DD-109), using a B480 grating with 2x900s dithered exposures and a central wavelength of 6250 \AA.
The data were reduced with \gls{dragons} \citep{labrieDRAGONSDataReduction2019}, and flexure corrections were applied by re-running the reduction software without sky subtraction, and calculating a scalar offset by comparing to a sky line catalog \citep{rousselotNightskySpectralAtlas2000}.
 An interesting feature observed in the Keck spectrum is some degree of asymmetry (skewed towards the redder wavelengths) in both the H$\alpha$ and MgII broad lines, as seen in Figure~\ref{fig:asymmetry}; however, the H$\alpha$ asymmetry does not appear in the later Gemini North-\gls{gmos} spectrum.
 The latter spectrum does not cover the MgII feature, and so no comparison between the two times is possible in this regime.
 Given that there is no similarly-resolved archival spectrum for this source, we are not able to distinguish transient spectral features from persistent host features at this time.
 
 We fit the Gemini spectrum using the Penalized Pixel-Fitting software (\texttt{ppxf}; \citealt{2023MNRAS.526.3273C}) in order to study the possible presence of an \gls{agn}.
 We fit the spectrum with stellar templates from the E-MILES SSP library \citep{2016MNRAS.463.3409V} along with a number of narrow and broad Gaussian peaks corresponding to various emission and absorption lines. Using these measured gas line strengths, we place this object in the \gls{bpt} diagram, finding that it lands in the ``composite'' region (according to the delineations of \citealt{2001ApJ...556..121K, 2003MNRAS.346.1055K, 2006MNRAS.372..961K}).
 This indicates that it is likely that the observed lines are due to a combination of star formation and \gls{agn} activity.
 We further measure the mass of the \gls{agn} using the fitted H$\alpha$ velocity dispersion, which we find to be $2282\pm 41$ km/s. The resulting \gls{bh} mass according to the method presented in \citep{2005ApJ...630..122G} is $M_{\rm SMBH}\sim 2\times10^7~M_\odot$.
 Note that this measurement may be inaccurate if the transient is significantly contributing to the \gls{blr} emission.

\gls{wise} \citep{wrightWIDEFIELDINFRAREDSURVEY2010} infrared data for the host of this event is shown in Figure~\ref{fig:C202309242206400m275139}.
Following \citet{Clark2024}, we retrieved this data from the NASA/IPAC infrared science archive (IRSA)\footnote{\url{https://irsa.ipac.caltech.edu/}}, utilizing data from both the ALLWISE \citep{wright_2010_WIDEFIELDINFRAREDSURVEY} and NEOWISE Reactivation Releases \citep[NEOWISE-R;][]{mainzer_2011_NEOWISEOBSERVATIONSNEAREARTH, mainzer_2014_INITIALPERFORMANCENEOWISE}.
Individual observations are combined following quality control cuts to obtain one mean magnitude per filter for each observation period, providing a $\sim$~6 month cadence.
\response{
The host is found to have brightened by $\sim$0.2 mag (with typical photometric uncertainties $\sim$0.05 mag) in both W1 and W2 bands starting around MJD 57900.
The SkyMapper data release \citep{skymapper} shows that the object brightened by $\sim$0.13/0.1 mag (uncertainties $\sim$0.05 mag) in $g$/$r$-band 5" aperture photometry over MJD 57870-57875.
A possible explanation for these phenomena identifies the \gls{ir} component as a reprocessed signature of the optical event; while blue, this event is slower than AT 2023aagj, which at one point brightened by $\sim$0.4 mag in both $g$ and $i$ band over a 5 day interval.
We also examine archival DECam and PanSTARRS \citep{Flewelling_2020} data covering this object, but are unable to put further constraints on past activity in part due to the sparse temporal coverage of these datasets in this region.
}

We also produce a \gls{wise} color-color plot for W1-W2 versus W2-W3, although this is not shown as W3 is only available in one epoch, which may not be representative of the entire color evolution of the object over the past years.
By comparison with populations of extragalactic objects (as defined in \citealt{wrightWIDEFIELDINFRAREDSURVEY2010}) we find that while the host most clearly lies in the \gls{wise} population regions associated with spiral and luminous infrared galaxies, it is near the boundary into the Seyfert and quasar parameter spaces, and 
has been observed to temporarily visit these regions.


We carried out a \gls{too} observation of AT 2023aagj with the \textit{Neil Gehrels Swift Observatory} \citep{gehrelsSwiftGammaRayBurst2004} X-ray Telescope \citep[XRT;][]{burrowsSwiftXRayTelescope2005}.
Our observation began on 2023-12-25 at 06:02 UT with a total exposure of 3265 s in \gls{pc} mode. 
We used the \textit{Swift}-XRT automated tools \citep{Evans2009,Evans2023} to analyze our data and all previous data covering the source position from May 2006 and April 2021 (totaling 7.8 ks exposure in \gls{pc} mode, including the latest data). 
We measure a source position of RA, DEC (J2000) = $22^{h}06^m 40^{s}.26$, $-27^\circ 51\arcmin 41.8\arcsec$ with an uncertainty radius of $4.7\arcsec$ (90\% confidence). 
In our latest observation the source displays a soft spectrum with all photon counts below 5 keV, similar to the previous observations.  
The time-averaged spectra were fit with an absorbed powerlaw model using \texttt{XSPEC v12.14.0} \citep{Arnaud1996} within \texttt{HEASoft v6.32}. 
The inferred photon index is relatively unconstrained but favors $\Gamma\approx2.5$ with hydrogen column density $N_H\approx2\times 10^{20}$ cm$^{-2}$. 
The unabsorbed flux ($0.3-10$ keV) is $F_X=(3.2^{+8.8}_{-1.0})\times10^{-13}$ erg cm$^{-2}$ s$^{-1}$ in our latest observation. 
We find a factor of $\sim$\,$4$ increase in flux between the 2006 and 2021 observations, but no significant change (a factor of $\sim$\,$1.2\pm0.5$) between 2021 and 2023 (due also to the larger errors on the flux). 
We therefore conclude that there is no ongoing X-ray outburst from AT 2023aagj, although we note that these observations were taken $\sim 2$ months after the optical peak at a time when any additional flux from the transient may have been too faint to be detected.
It is interesting to note how the \gls{wise} flare peaked around 2018, notably coincident with the significant change in X-ray flux. 
We speculate this may be related to a changing look \gls{agn} event, although this is hard to confirm due to the lack of an older spectrum.

\subsection{Other candidates of scrutiny}
\label{subsec:othercands}

We took spectra for three additional candidates beyond AT 2023aagj during our follow-up campaign.
The spectrum for one transient (AT 2023unl) provides inconclusive evidence for the nature of the source; however, the source was observed to redden in later epochs, and so is disfavored by our assumed counterpart model.
The spectra for the other two transients (AT 2023aden and AT 2023uab) contain \gls{sn}-like features; both of these transients are subsequently excluded as possible counterparts to \gweventid.
The three spectra are shown in Figure~\ref{fig:plot_spectra_other} (Appendix \ref{app:additonaldata}).

\subsubsection{AT 2023unl (C202310042207549m253435)}
\label{subsubsec:AT2023unl}

The spectrum for AT 2023unl was taken by the \gls{rss} \citep{burghPrimeFocusImaging2003, kobulnickyPrimeFocusImaging2003} on the \gls{salt} \citep{buckleyCompletionCommissioningSouthern2006}.
The \texttt{RSSMOSPipeline} \citep{2018ApJS..235...20H}\footnote{\url{https://github.com/mattyowl/RSSMOSPipeline}} was used for the reduction of \gls{salt}-\gls{rss} data.
The pipeline automatically identifies and extracts one dimensional spectra using the data products delivered by the \gls{salt} observation team.
Using the known redshift of the observed object and the reduced spectra, \texttt{PYQSOFit} \citep{2018ascl.soft09008G}\footnote{\url{https://github.com/legolason/PyQSOFit}} was then used to fit and identify possible emission lines in the data.
The \gls{snr} for the spectrum is relatively low, with few strong features.
The present H$\beta$ line does not demonstrate any unusual characteristics such as asymmetry.
A possible broad line feature is present around an observed wavelength of 5525 \AA, albeit a chip gap appears to have ended up at the peak of the feature, and so it is uncertain how much of the observed variation is due to imperfect data reduction; that said, there is no similar feature around the second chip gap near 6575 \AA. We conclude there is no conclusive evidence for this transient being in an \gls{agn}.

\subsubsection{AT 2023aden (A202310262246341m291842)}
\label{subsubsec:AT2023aden}

A spectrum was taken for this transient using Gemini North \gls{gmos} on November 16, 2023 (PI: Cabrera, PID: GN-2023B-DD-103), and was acquired and reduced in a similar manner as the Gemini spectrum for AT 2023aagj, with a central wavelength of 7600 \AA.
The spectrum is blue and has strong absorption features characteristic of \glspl{sn}; subsequently, we do not consider this transient as a likely counterpart to \gweventid.

\subsubsection{AT 2023uab (C202309242248405m134956)}
\label{subsubsec:AT2023uab}

We used the Double Spectrograph (DBSP) mounted at the 200-inch Hale Telescope at Palomar Observatory.
For this source, we used a 1.5 arcsec slitmasks, a D55 dichroic, a blue grating of 600/4000 and red grating of 316/7500.
The data were reduced using a custom PyRAF DBSP reduction pipeline \citep{dbsp}. 
The spectrum exhibits clear H$\alpha$ and H$\beta$ features which enable a redshift measurement of 0.128 (computed with \texttt{Redrock}\footnote{\url{https://github.com/desihub/redrock}}; Bailey et al., in prep.).
The continuum of the spectrum peaks around $\sim$5000 \AA, with several broad absorption features present around the peak.
This, in concert with \gls{parsnip} classifying the object as a type Ia \gls{sn}, leads us to exclude this object as a possible counterpart.

\section{Discussion}
\label{sec:discussion}

\response{
We return to our favored counterpart candidate AT 2023aagj and assess its significance as an \gls{agn} transient.
The persistent blue color of the transient distinguishes it from the phenomenology of typical supernovae, a separation further supported by a \gls{parsnip} classification of ``TDE"/``Non-SN" as noted above.
We apply \gls{sf} arguments as presented in \citep{palmeseLIGOVirgoBlack2021} to the photometry to calculate the probability that the observed variation occurred as a result of typical \gls{agn} activity.
We find the variability of AT 2023aagj significant on the 5.9$\sigma$ level, with an associated one-sided Gaussian probability of $\sim 10^{-16}$; such a low value makes it highly unlikely that AT 2023aagj is a feature of the host \gls{agn}'s activity, and strongly prefers the model of AT 2023aagj as a specially transient phenomenon.
}
Taking the 90\% \gls{ci} volume of the event as $3.73 \times 10^8$ Mpc$^3$ (as calculated with \texttt{ligo.skymap}) and fiducial \gls{agn} number densities of $10^{-4.75}$ Mpc$^{-3}$ \response{and $10^{-4}$ Mpc$^{-3}$} (\citealt{bartosGravitationalwaveLocalizationAlone2017a,Greene:2007xw}; \response{the two values reflect expected and conservative (more \gls{agn}/background flares) rates}), we estimate counts of 6600 \response{and 37000 \gls{agn}} in the volume.
Alternatively, crossmatching the Milliquas \citep{fleschMillionQuasarsMilliquas2023} quasar catalog with the skymap (also with \texttt{ligo.skymap}) yields a 3D match of 826 objects.
Given the discovery window of our follow-up campaign $t_{\rm window} = 92$ days (calculated as the difference in time between the second epoch and the last), we calculate the probability of a chance discovery of a similar flare as
\begin{equation}
    p_{cc}
    = n_{\rm AGN} \frac{t_{\rm window}}{\Delta t_{\rm rise}} p_{\rm flare}
    \approx \left( 4.48 \times 10^{-16} \frac{\text{flares}}{\text{AGN}} \right) n_{\rm AGN},
\end{equation}
which predicts $p_{cc} \sim \mathcal{O}(10^{-12})$ \response{and $\mathcal{O}(10^{-11})$}, and $p_{cc} \sim \mathcal{O}(10^{-13})$ for the number density and Milliquas estimates, respectively.
\response{
We note that while these rates distinguish AT 2023aagj as a special event, due to uncertainties in the understanding of \gls{bbh} counterpart morphology the identity of AT 2023aagj as a counterpart to S230922g remains largely uncertain. A caveat of this analysis is that we also cannot account for the variability of this specific AGN given the lack of long-term monitoring from archival data.}

We compare our candidate flare with the sample of \gls{ztf} transients from \citet{grahamLightDarkSearching2023}.
Assuming a roughly constant spectral flux density across the $g$- and $B$-band frequency domain, we apply the $B$ bandpass bolometric correction of 5.15 from \citet{durasUniversalBolometricCorrections2020} to derive a total flare energy of
\begin{equation}
\begin{split}
    E_{\rm tot} &
    \approx
    4 \pi d_L^2
    \left( 5.15 \int \mathrm{d}\nu T_B (\nu) \right)
    \int_{t_d}^{t_e} \mathrm{d}t 10^{-0.1 m_{g,\rm AB}(t) + 48.6} \\
    & \approx
    1.2 \times 10^{50} \text{ erg},
\end{split}
\end{equation}
where $T_B(\nu)$ is the transmission function for the $B$ bandpass (\citealt{bessellUBVRIPassbands1990}),  and $t_d$ and $t_e$ are the times of the first and last observations of the flare.
Fitting a Gaussian rise-exponential decay model to the $g$-band light curve yields a rise time parameter (the standard deviation of the Gaussian) of $t_{\rm rise} \approx 5.25 \text{ days}$.
This places this transient in an intermediate regime between the \gls{sn} and \gls{tde} populations from \citealt{grahamLightDarkSearching2023} (c.f. Figure 2 in that paper) among fast \glspl{tde} and high-energy \glspl{sn}.
Notably, the \citealt{grahamLightDarkSearching2023} proposed flare counterparts are all slower and more energetic than AT 2023aagj; this may be associated with the measurement of our flare parameters being based on difference photometry, while the \gls{ztf} light curves used in the compared study are assembled via direct photometry that includes the flux of the host \gls{agn}.

We estimate physical parameters of the system following the methodology of \citet{grahamCandidateElectromagneticCounterpart2020}, which tests the hypothesis that the flare is generated 
from a Bondi accreting \gls{bh} as the remnant emerges from the accretion disk due to a post-merger kick.
With a flare onset time delay of $\sim2$ days from the \gls{gw} trigger and total mass estimate of $M_{\rm tot} \sim 90 M_\odot$, under the assumption of association we can conclude one or more of the following: either (i) the kick velocity $v_{\rm kick}$ is large, or (ii) the disk is geometrically thin (aspect ratio $h \leq \mathcal{O}(10^{-3})$) or (iii) the disk is not optically thick away from a geometrically thin mid-plane, or (iv) there is a cavity in the disk due to feedback from the pre-merger \gls{bbh}.

We continue by assuming the kick velocity is not specially large or small.
If we assume a kick velocity $v_{\rm kick} \sim \mathcal{O}(200){\rm km/s}$ (which is consistent with the peak of the prior in \citealt{Varma22}) then we can constrain the approximate disk height from \citep{grahamLightDarkSearching2023}
\begin{equation}
    t_{\rm exit} \sim \frac{H \sqrt{2 {\rm ln} \tau_{\rm mp}}}{v_{\rm kick}}
\end{equation}
where $H$ is the disk height and $\tau_{\rm mp}$ is the mid-plane optical depth.
Since $t_{\rm exit} \sim 2\text{ days}$, $\tau_{\rm mp} \sim 10^{[3,6]}$ yields a factor $4-5$ from the square root, so $H \sim 0.3r_{g}(M_{\rm SMBH}/2\times 10^{7}M_{\odot})$.
This implies a thin disk, possibly similar to the thin regions of a \citet{TQM05} model, which also has a lower $\tau_{\rm mp}$ than other models such as \citet{SG03}.
The thinner regions of \gls{agn} disks are where gas damping is most efficient and consequently where mergers are more likely to occur \citep{McK24}.

If we assume from Figure~\ref{fig:C202309242206400m275139} that the flare begins 2 days post-\gweventid, and the flare lasts through the spectra denoted by the blue and red lines, then the overall flare duration is $t_{\rm flare} \sim \mathcal{O}(10^{2})$days. \response{We note that this is to be considered a rough estimate of the flare timescale in what follows.}
Given $E_{\rm tot} \sim 10^{50}{\rm erg}$, the average luminosity of the flare is ${\overline{L}_{\rm flare}} \sim 10^{43}{\rm erg/s}$, which can be parameterized as $\overline{L}_{\rm flare}  \sim 10^{3}L_{\rm Edd}(M_{\rm BBH}/90M_{\odot})$ where $L_{\rm Edd}$ is the Eddington luminosity.
If we parameterize the Bondi accretion rate onto the merged \gls{bh} ($\dot{M}_{\rm BHL}$) as:
\begin{multline}
\dot{M}_{\rm BHL} \sim \frac{0.03M_{\odot}}{{\rm yr}} \left(\frac{M_{\rm BBH}}{90M_{\odot}} \right)^{2} \\
\times \left(\frac{v_{\rm rel}}{200{\rm km/s}}\right)^{-3} \left(\frac{\rho_{\rm disk}}{10^{-11}{\rm g/cm^{3}}} \right)
\end{multline}
where $\rho_{\rm disk}$ is the disk gas density and we assume the relative velocity ($v_{\rm rel}$) between the \gls{bh} and the \gls{agn} disk gas is $v_{\rm rel} \sim v_{\rm kick}$, i.e. the sound speed in the \gls{agn} disk gas is less than $v_{\rm kick}$.
We have chosen $\rho_{\rm disk}$ to be comparable to the typical density in a \citet{TQM05} model where the disk is near its thinnest ( aspect ratio $h \sim 10^{-3}$). Then $ L_{\rm flare}\sim 10^{-2}L_{\rm BHL}$, where
$L_{\rm BHL} = \eta \dot{M}_{\rm BHL} c^{2}$.
Interestingly, this accretion rate is within an order of magnitude of the inferred accretion rate onto embedded \gls{bbh} in \gls{agn} from recent GRMHD simulations \citep{Dittmann24}.

Associating the line asymmetry with an off-center flare implies that the signature persists in the \gls{blr} for $\sim 70\text{ days}$ post event, but is gone $\sim 90\text{ days}$ post event.
A light-travel time across the full \gls{blr} of $\mathcal{O}(90\text{ days})$ ($\sim 80$ in rest frame), corresponds to a distance scale of $0.07{\rm pc} \sim 7 \times 10^{4}r_{g}(M_{\rm SMBH}/2 \times 10^{7}M_{\odot})$, implying that the merger occurred around $4 \times 10^{4}r_{g}(M_{\rm SMBH}/2 \times 10^{7}M_{\odot})$.
This is located further out than the thinnest regions of a \citet{SG03} model disk, so we can rule out an origin in a disk similar to that model. This location could be consistent with the thinnest ($h \sim 10^{-3}$) regions of a \citet{TQM05} model disk, which extend $\mathcal{O}[0.01,0.1]{\rm pc}$.

An alternative candidate for an off-center luminous flare in an \gls{agn} is a $\mu$-\gls{tde}, where a star is tidally disrupted by a stellar mass \gls{bh} embedded in an \gls{agn} disk.
\citet{Perna21} show that in the outer regions ($\sim 7 \times 10^{4}r_{g}(M_{\rm SMBH}/2 \times 10^{7}M_{\odot}$) of a \citet{TQM05} disk model, a $\mu$-\gls{tde} will peak in afterglow very quickly ($\sim$hours), which is inconsistent with our observations.
However, if $M_{\rm SMBH}$ is actually a factor of a few more massive, then the diffusion time becomes longer, potentially months, which could be consistent with our observations.

We also note that the lightcurve of AT 2023aagj resembles those of afterglows in \gls{agn} disks as found in \cite{Wang_2022}.
The rise of an afterglow from a \gls{bbh} merger is not implausible, assuming that these objects are able to launch jets, as explored by \citet{tagawaObservableSignatureMerging2023}.
Future work could investigate the possibility of fitting those afterglow models in high-density environments.

Under the assumption that AT 2023aagj is the counterpart to S230922g, our requirements are that the \gls{agn} disk must be relatively thin, similar to a \citet{TQM05} model, and that the accretion rate onto the kicked \gls{bh} is significantly super-Eddington, but well below the Bondi rate (and requires an associated jet for the radiation to emerge).
One of our most likely false positive for this event is a $\mu$-\gls{tde} in the disk if our estimate of $M_{\rm SMBH}$ is too small by a factor of a few.




\section{Conclusions}
\label{sec:conclusions}

In this work we present the results of the \gls{gwmmads} follow-up of \gweventid.
We discuss in detail our most likely candidate AT 2023aagj, noting it is mostly constant blue color, \gls{agn}-hosted nature, and especially the presence of asymmetry in \gls{blr} spectral features as evidence in favor of recognizing it as a counterpart to \gweventid.
However, we do not find the current dataset sufficient to clearly identify the transient as such.

It is worth noting that several aspects lead to an inconclusive outcome regarding the significance of this association.
The variable nature of the \gls{agn} in question, showing a significant flare in the infrared and a change in X-ray flux over the past decade, exemplifies that there is some inherent activity in the host, and the discovery of a \gls{gw} counterpart is only possible if a means of distinguishing the transient signal from any host variability.
One possible discriminator is the observation of asymmetric components in the broad lines of the \gls{agn} spectrum, indicative of an off-center phenomenon; however, other sources such as disk winds and tilted dust obscuration may cause similar profiles.
A smoking gun for the \gls{bbh} association would be the evolution of the asymmetric component in sync with the transient light curve, with the asymmetry shifting towards bluer wavelengths and eventually disappearing as the transient fades.
For AT 2023aagj, due to the lack of wavelength coverage for the Gemini spectrum and subsequent lack of multiple observations of MgII, a chip gap on H$\beta$ in the Keck spectrum, the telluric region blueward of H$\alpha$, and the target setting in December, the extraction of such an evolution is not feasible with the present data.
A spectrum of this object in 2024 as the target rises again may be informative about the nature of the asymmetry.
A persistent line asymmetry long after the merger event may point to similarly persistent asymmetric illumination in this source, such as a warped disk, and could rule out the association between AT 2023aagj and \gweventid.
Finally, given the blue color and the \gls{parsnip} prediction, we cannot exclude that this transient may have originated from a \gls{tde}.

We consider the uncertainty of our result as another indicator of the challenging nature of the search for \gls{bbh} \gls{em} counterparts and successively the need for the continuation and growth of future efforts for this purpose.
More so than for \gls{ns} merger counterparts, predictions of \gls{bbh} counterparts share a parameter space with existing transient families (e.g. that of \glspl{tde}) and AGN variability, and distinguishing between the different classes of events at the present time may be challenging.
Spectroscopic data are expected to be the most helpful towards this end, and so it is important to combine the appropriate resources with photometric follow-up while transients are still active, for as long as it takes to develop models mature enough to perform well without requiring such expensive resources.
This is especially relevant for the remainder of \gls{o4} follow-up, as all efforts made now will be the last \say{uptime} for \gls{gw} follow-up until \gls{o5} begins in 2027.
With Virgo having joined the \gls{gw} detector network for \gls{o4}b, we expect future \gls{o4} searches to yield better localized events, a trend that is expected to continue with \gls{o5}; these lower search volumes, in concert with forthcoming powerful resources such as \gls{lsst} \citep{ivezicLSSTScienceDrivers2019}, will help future counterpart searches be more efficient and effective, and will help deepen and broaden the impact of \gls{gw} and multimessenger astronomy in the years to come.

\section*{Data and Code Availability}

\response{
The data products and figure code for this publication are available at \url{https://doi.org/10.5281/zenodo.13787730}.
}

\section*{Acknowledgements}

TC, AP, and LH acknowledge that this material is based upon work supported by NSF Grant No. 2308193. BO gratefully acknowledges support from the McWilliams Postdoctoral Fellowship at Carnegie Mellon University. BM \& KESF are supported by NSF AST-2206096 and NSF AST-1831415 and Simons Foundation Grant 533845 as well as Simons Foundation sabbatical support. The Flatiron Institute is supported by the Simons Foundation. AP thanks Rosalba Perna, Armin Rest, and Stephen Smartt for useful discussion. This work was supported by the Deutsche Forschungsgemeinschaft (DFG, German Research Foundation) under Germany's Excellence Strategy – EXC-2094 – 390783311.

This research used resources of the National Energy Research Scientific Computing Center, a DOE Office of Science User Facility supported by the Office of Science of the U.S.
Department of Energy under Contract No. DE-AC02-05CH11231 using NERSC award HEP-ERCAP0029208 and HEP-ERCAP0022871. This work used resources on the Vera Cluster at the Pittsburgh Supercomputing Center (PSC).
We thank T.J. Olesky and the PSC staff for help with setting up our software on the Vera Cluster.
MWC acknowledges support from the National Science Foundation with grant numbers PHY-2308862 and PHY-2117997.

This project used data obtained with the Dark Energy Camera (DECam), which was constructed by the Dark Energy Survey (DES) collaboration.
Funding for the DES Projects has been provided by the US Department of Energy, the US National Science Foundation, the Ministry of Science and Education of Spain, the Science and Technology Facilities Council of the United Kingdom, the Higher Education Funding Council for England, the National Center for Supercomputing Applications at the University of Illinois at Urbana-Champaign, the Kavli Institute for Cosmological Physics at the University of Chicago, Center for Cosmology and Astro-Particle Physics at the Ohio State University, the Mitchell Institute for Fundamental Physics and Astronomy at Texas A\&M University, Financiadora de Estudos e Projetos, Fundação Carlos Chagas Filho de Amparo à Pesquisa do Estado do Rio de Janeiro, Conselho Nacional de Desenvolvimento Científico e Tecnológico and the Ministério da Ciência, Tecnologia e Inovação, the Deutsche Forschungsgemeinschaft and the Collaborating Institutions in the Dark Energy Survey.

The Collaborating Institutions are Argonne National Laboratory, the University of California at Santa Cruz, the University of Cambridge, Centro de Investigaciones En\`ergeticas, Medioambientales y Tecnol\`ogicas–Madrid, the University of Chicago, University College London, the DES-Brazil Consortium, the University of Edinburgh, the Eidgenössische Technische Hochschule (ETH) Zürich, Fermi National Accelerator Laboratory, the University of Illinois at Urbana-Champaign, the Institut de Ci\'encies de l’Espai (IEEC/CSIC), the Institut de F\'isica d’Altes Energies, Lawrence Berkeley National Laboratory, the Ludwig-Maximilians Universit\:at M\:unchen and the associated Excellence Cluster Universe, the University of Michigan, NSF’s NOIRLab, the University of Nottingham, the Ohio State University, the OzDES Membership Consortium, the University of Pennsylvania, the University of Portsmouth, SLAC National Accelerator Laboratory, Stanford University, the University of Sussex, and Texas A\&M University.

Based on observations at Cerro Tololo Inter-American Observatory, NSF’s NOIRLab (NOIRLab Prop. ID 2022B-715089; PI: Palmese; 2023B-851374 , PI: Andreoni \& Palmese; 2023B-735801, PI: Palmese \& Wang), which is managed by the Association of Universities for Research in Astronomy (AURA) under a cooperative agreement with the National Science Foundation.
We thank Kathy Vivas, Alfredo Zenteno, and CTIO staff for their support with DECam observations.

Based on observations obtained at the international Gemini Observatory (Prop. ID GN-2023B-DD-103, PI: Cabrera; GN-2023B-109, PI: Cabrera), a program of NSF NOIRLab, which is managed by the Association of Universities for Research in Astronomy (AURA) under a cooperative agreement with the U.S. National Science Foundation on behalf of the Gemini Observatory partnership: the U.S. National Science Foundation (United States), National Research Council (Canada), Agencia Nacional de Investigaci\'{o}n y Desarrollo (Chile), Ministerio de Ciencia, Tecnolog\'{i}a e Innovaci\'{o}n (Argentina), Minist\'{e}rio da Ci\^{e}ncia, Tecnologia, Inova\c{c}\~{o}es e Comunica\c{c}\~{o}es (Brazil), and Korea Astronomy and Space Science Institute (Republic of Korea).
Gemini data were processed using DRAGONS (Data Reduction for Astronomy from Gemini Observatory North and South).

This work was enabled by observations made from the Gemini North telescope, located within the Maunakea Science Reserve and adjacent to the summit of Maunakea. We are grateful for the privilege of observing the Universe from a place that is unique in both its astronomical quality and its cultural significance.

Some of the observations reported in this paper were obtained with the Southern African Large Telescope (SALT).

The Legacy Surveys consist of three individual and complementary projects: the Dark Energy Camera Legacy Survey (DECaLS; Proposal ID 2014B-0404; PIs: David Schlegel and Arjun Dey), the Beijing-Arizona Sky Survey (BASS; NOAO Prop. ID 2015A-0801; PIs: Zhou Xu and Xiaohui Fan), and the Mayall z-band Legacy Survey (MzLS; Prop. ID 2016A-0453; PI: Arjun Dey). DECaLS, BASS and MzLS together include data obtained, respectively, at the Blanco telescope, Cerro Tololo Inter-American Observatory, NSF’s NOIRLab; the Bok telescope, Steward Observatory, University of Arizona; and the Mayall telescope, Kitt Peak National Observatory, NOIRLab. Pipeline processing and analyses of the data were supported by NOIRLab and the Lawrence Berkeley National Laboratory (LBNL). The Legacy Surveys project is honored to be permitted to conduct astronomical research on Iolkam Du’ag (Kitt Peak), a mountain with particular significance to the Tohono O’odham Nation. LBNL is managed by the Regents of the University of California under contract to the U.S. Department of Energy.

BASS is a key project of the Telescope Access Program (TAP), which has been funded by the National Astronomical Observatories of China, the Chinese Academy of Sciences (the Strategic Priority Research Program “The Emergence of Cosmological Structures” Grant XDB09000000), and the Special Fund for Astronomy from the Ministry of Finance. The BASS is also supported by the External Cooperation Program of Chinese Academy of Sciences (Grant  114A11KYSB20160057), and Chinese National Natural Science Foundation (Grant  12120101003, 11433005).

The Legacy Survey team makes use of data products from the Near-Earth Object Wide-field Infrared Survey Explorer (NEOWISE), which is a project of the Jet Propulsion Laboratory/California Institute of Technology. NEOWISE is funded by the National Aeronautics and Space Administration.

The Legacy Surveys imaging of the DESI footprint is supported by the Director, Office of Science, Office of High Energy Physics of the U.S. Department of Energy under Contract No. DE-AC02-05CH1123, by the National Energy Research Scientific Computing Center, a DOE Office of Science User Facility under the same contract; and by the U.S. National Science Foundation, Division of Astronomical Sciences under Contract No. AST-0950945 to NOAO.

This research has made use of the NASA/IPAC Extragalactic Database (NED), which is funded by the National Aeronautics and Space Administration and operated by the California Institute of Technology.

\facilities{Blanco (DECam), Keck:I (LRIS), Gemini:South (GMOS), SALT (RSS), Hale (DBSP), WO:2m (3KK)}

\software{
astropy \citep{astropycollaborationAstropyCommunityPython2013, astropycollaborationAstropyProjectBuilding2018, astropycollaborationAstropyProjectSustaining2022},
DRAGONS \citep{labrieDRAGONSDataReduction2019},
dustmaps \citep{greenDustmapsPythonInterface2018},
gwemopt \citep{coughlinOptimizingSearchesElectromagnetic2018},
healpy \citep{gorskiHEALPixFrameworkHighResolution2005, zoncaHealpyEqualArea2019},
ppxf \citep{2023MNRAS.526.3273C},
Fritz SkyPortal \citep{coughlinDataSciencePlatform2023},
ligo.skymap \citep{singerGoingDistanceMapping2016},
LPIPE \citep{lpipe},
matplotlib \citep{hunterMatplotlib2DGraphics2007},
numpy \citep{harrisArrayProgrammingNumPy2020},
pandas \citep{mckinneyDataStructuresStatistical2010},
ParSNIP \citep{booneParSNIPGenerativeModels2021},
pyraf-dbsp \citep{dbsp},
Redrock (\url{https://github.com/desihub/redrock}),
SExtractor \citep{bertinSExtractorSoftwareSource1996a},
SFFT \citep{huImageSubtractionFourier2022},
SWarp \citep{bertinTERAPIXPipeline2002}
}

\appendix

\section{Additional data}
\label{app:additonaldata}

\begin{figure*}
    \centering

    \gridline{
    \fig{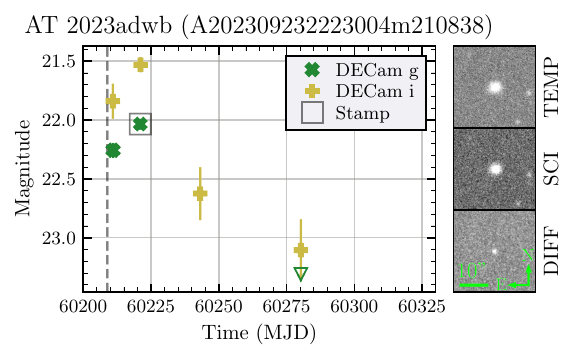}{0.49\textwidth}{}
    \fig{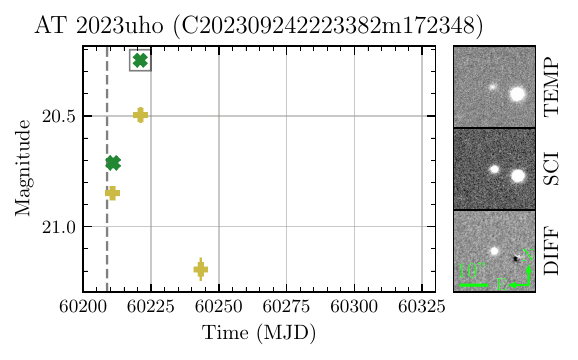}{0.49\textwidth}{}
    }
    \gridline{
    \fig{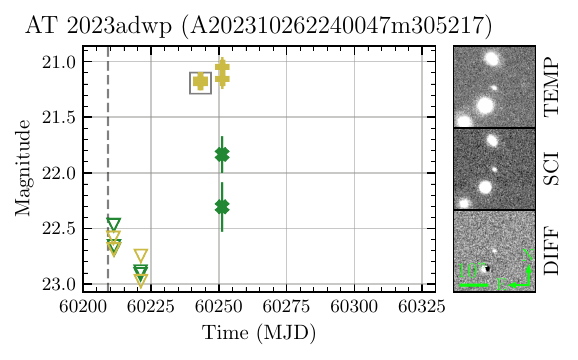}{0.49\textwidth}{}
    \fig{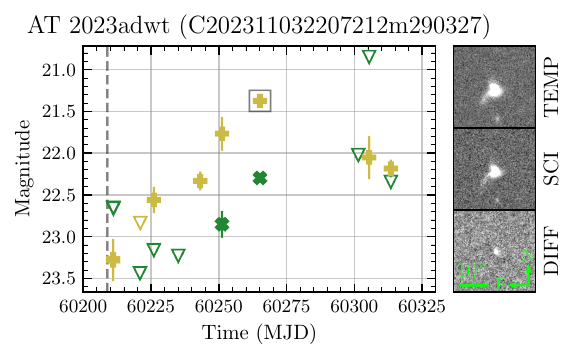}{0.49\textwidth}{}
    }
    \gridline{
    \fig{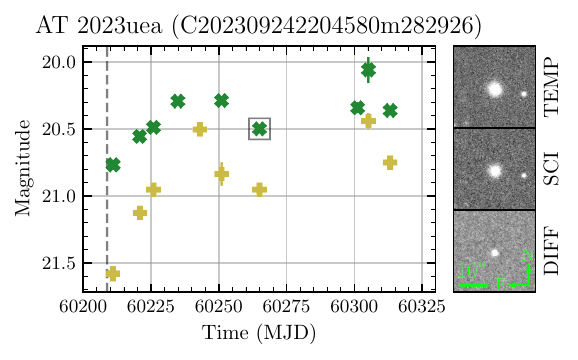}{0.49\textwidth}{}
    \fig{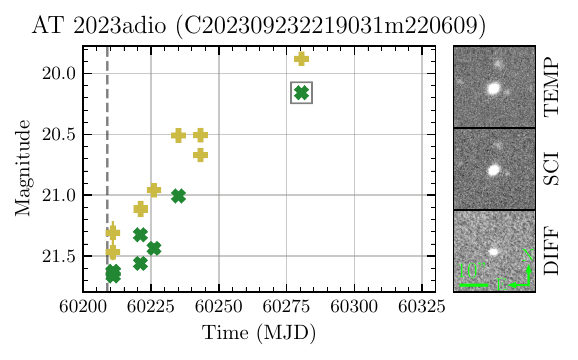}{0.49\textwidth}{}
    }
    \caption{
        Light curves for our remaining 22 candidates (continued in following figures).
        The dashed line indicates the S230922g event time.
        The sample stamps for each transient are taken from the exposure with the highest SNR, indicated with a gray square.
        Data taken with Wendelstein appear as small diamonds, where relevant (the red point for AT 2023uab is $J$-band).
    }
    \label{fig:light_curves_other}
\end{figure*}

\begin{figure*}
    \centering

    \gridline{
    \fig{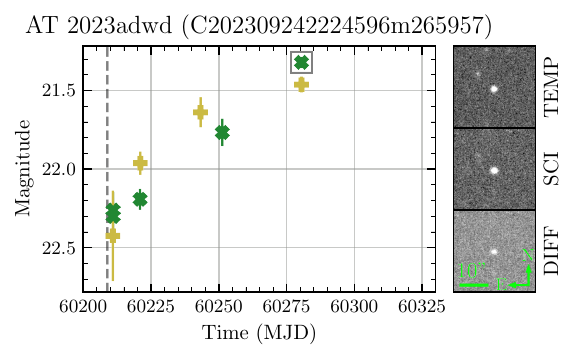}{0.49\textwidth}{}
    \fig{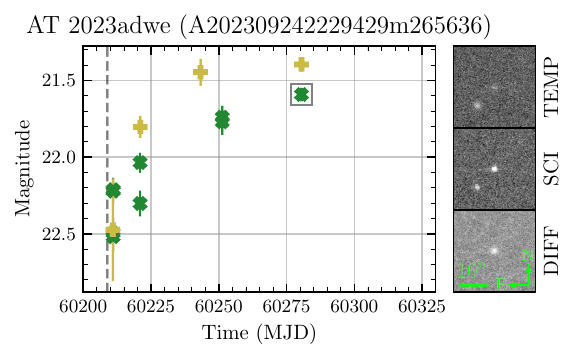}{0.49\textwidth}{}
    }
    \gridline{
    \fig{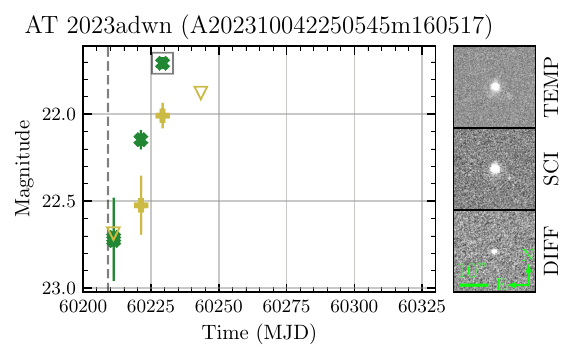}{0.49\textwidth}{}
    \fig{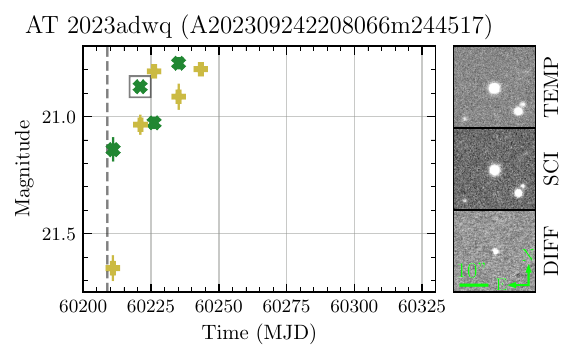}{0.49\textwidth}{}
    }
    \gridline{
    \fig{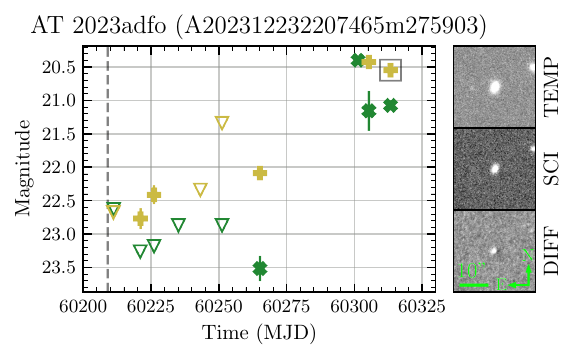}{0.49\textwidth}{}
    \fig{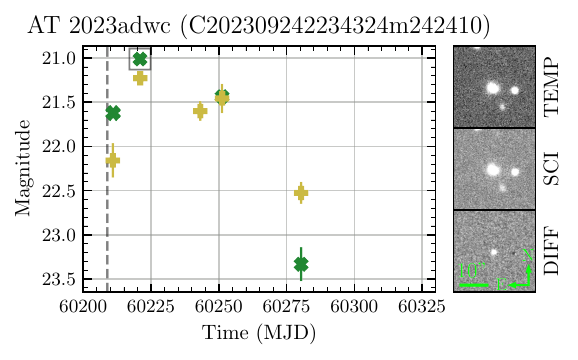}{0.49\textwidth}{}
    }
    \caption{Light curves for our remaining 22 candidates (cont.).}
    \label{fig:light_curves_other.1}
\end{figure*}

\begin{figure*}
    \centering

    \gridline{
    \fig{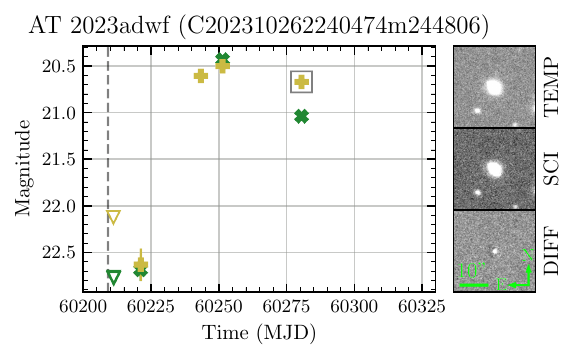}{0.49\textwidth}{}
    \fig{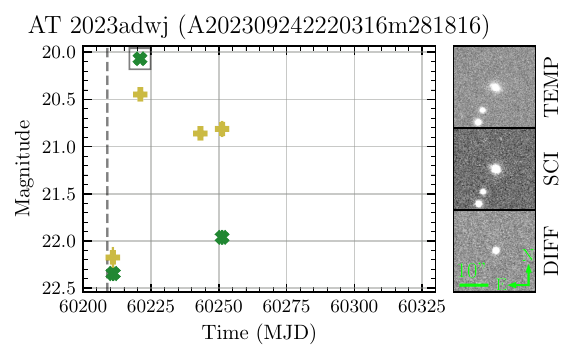}{0.49\textwidth}{}
    }
    \gridline{
    \fig{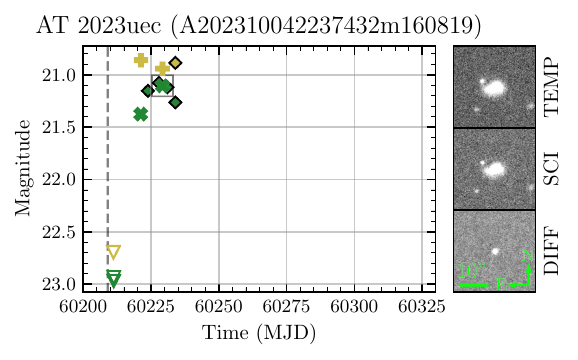}{0.49\textwidth}{}
    \fig{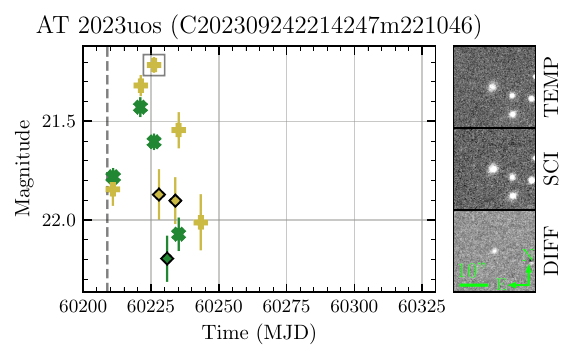}{0.49\textwidth}{}
    }
    \gridline{
    \fig{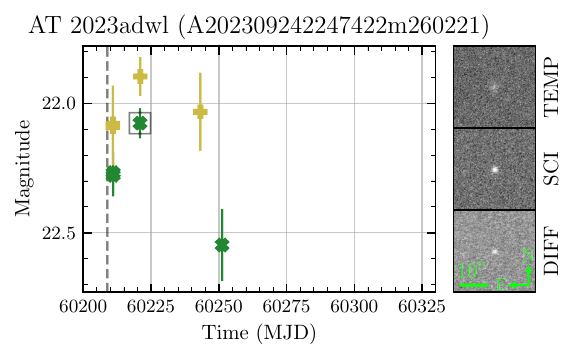}{0.49\textwidth}{}
    \fig{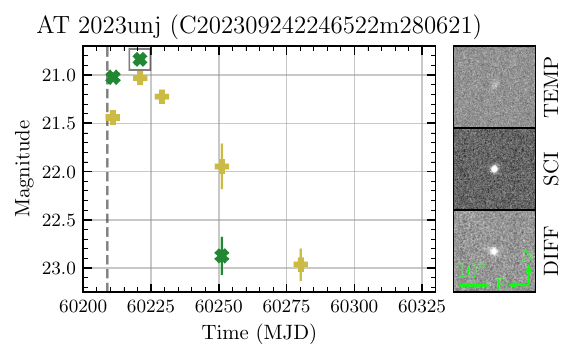}{0.49\textwidth}{}
    }
    \caption{Light curves for our remaining 22 candidates (cont.).}
    \label{fig:light_curves_other.2}
\end{figure*}

\begin{figure*}
    \centering

    \gridline{
    \fig{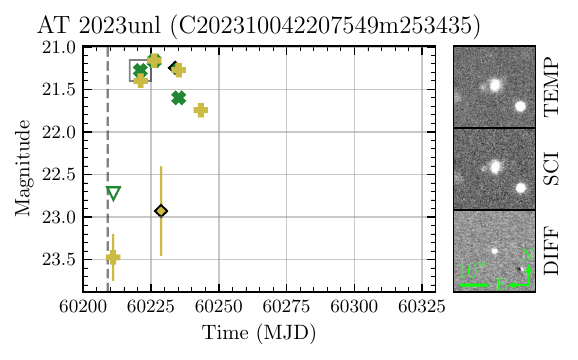}{0.49\textwidth}{}
    \fig{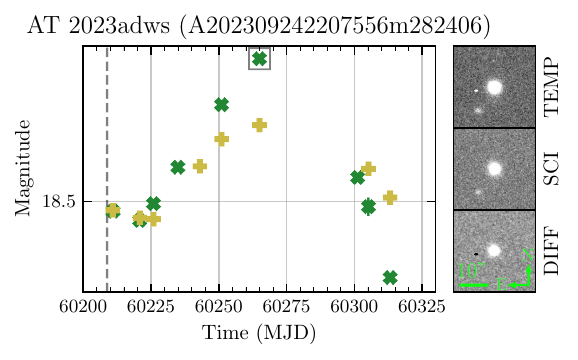}{0.49\textwidth}{}
    }
    \gridline{
    \fig{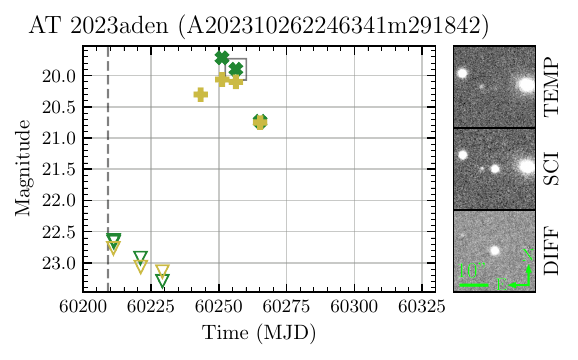}{0.49\textwidth}{}
    \fig{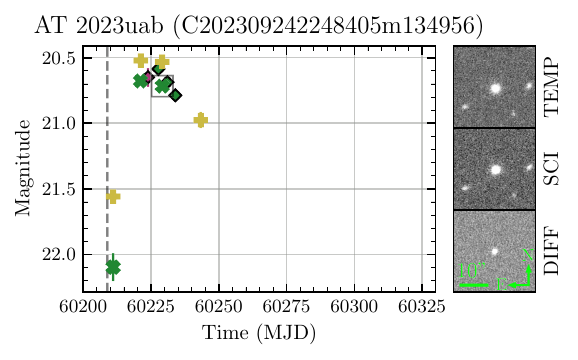}{0.49\textwidth}{}
    }
    \caption{Light curves for our remaining 22 candidates (cont.).}
    \label{fig:light_curves_other.3}
\end{figure*}

\begin{figure*}
    \centering

    \gridline{
    \fig{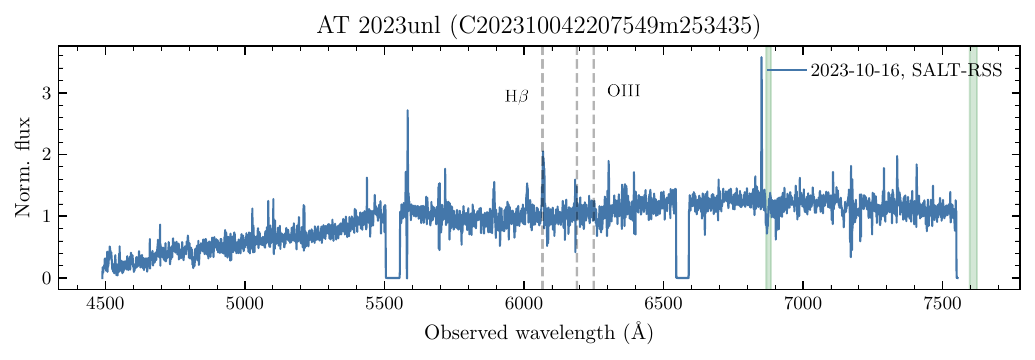}{0.95\textwidth}{}
    }
    \gridline{
    \fig{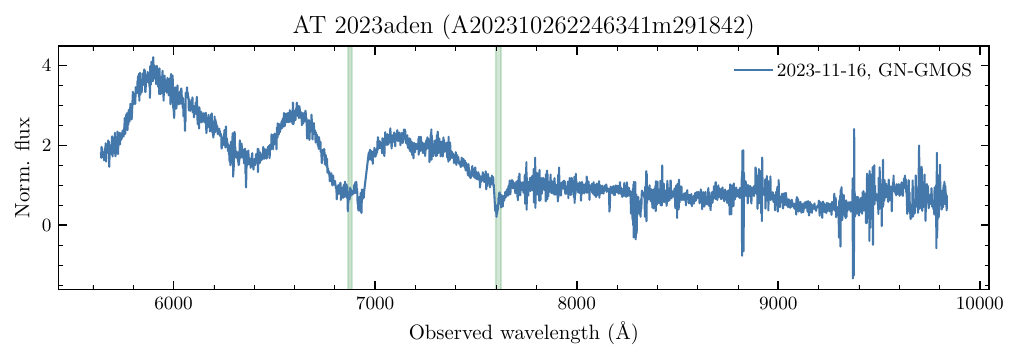}{0.95\textwidth}{}
    }
    \gridline{
    \fig{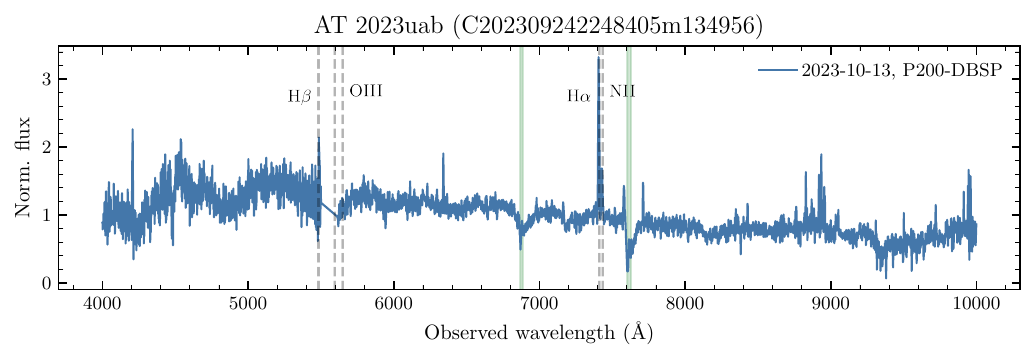}{0.95\textwidth}{}
    }
    \caption{
        Additional spectra taken as a part of our follow-up campaign.
        A selection of spectral lines and telluric features are marked as in Figure \ref{fig:C202309242206400m275139}.
    }
    \label{fig:plot_spectra_other}
\end{figure*}

This section contains light curves (Figures~\ref{fig:light_curves_other}-\ref{fig:light_curves_other.3}) and spectra (Figure~\ref{fig:plot_spectra_other}) for all candidates other than AT 2023aagj.

For some transients we collected additional data with the \gls{3kk} \citep{2016SPIE.9908E..44L} at the 2.1m-Fraunhofer Wendelstein Telescope \citep{2014SPIE.9145E..2DH} located in the German alps.
The instrument takes three images at the same time in two optical passbands and one \gls{nir} band.
For the monitoring of the transients we used the $g$, $i$ and $J$ bands.
For the detrending of the images the pipeline makes use of the tools from \cite{2002A&A...381.1095G}.
Then the images are calibrated and coadded using \texttt{SExtractor} \citep{1996A&AS..117..393B}, \texttt{SCAMP} \citep{2006ASPC..351..112B} and \texttt{SWarp} \citep{2002ASPC..281..228B}, difference photometry is conducted with \gls{sfft} \citep{huImageSubtractionFourier2022}, and extinction corrections are applied as described in \S\ref{subsec:autovet}.
During this process, the astrometric solution is computed against the \textit{Gaia} EDR3 catalog \citep{gaiacollaborationGaiaMission2016, 2021A&A...649A...1G} and the zero points for the optical bands are calibrated with the \glspl{ps1} catalog \citep{2013ApJS..205...20M}.
The \glspl{nir} zero points are matched to the \glspl{2mass} catalog \citep{2006AJ....131.1163S} and converted to AB magnitudes \citep{2007AJ....133..734B}.
Wendelstein data are plotted as outlined diamonds in Figure~\ref{fig:light_curves_other}, with $g$, $i$, and $J$ band data plotted in green, yellow, and red, respectively.


\bibliography{GW-MMADS_S230922g}

\end{document}

%% file: candidates_table.tex
\begin{deluxetable*}{cccccccc}
    \tablecaption{
        Summary table for our counterpart candidate shortlist.
        Redshifts are shown as available from crossmatching with several extragalactic databases and direct measurement from our spectra.
        The luminosity distances and uncertainties are reproduced from the GW skymap, using the \texttt{DISTMU} and \texttt{DISTSIGMA} values for the HEALPix tile in which the transient is located.
        The objects are sorted by ascending 2D skymap probability CI, s.t. the objects in the highest probability regions are listed first.
        The highest probability ParSNIP photometric classification along with the probability are listed in the last two columns; in this work, we rename the ParSNIP class ``TDE" as ``Non-SN" (see text).
        The last three subdivisions of the table include transients that did not peak during our observation window, those that reddened in later epochs, and those that were exluded as possible counterparts through spectroscopic classification.
        \label{tab:candidates}
    }
    \tablehead{
        \colhead{Object} & \multicolumn{2}{c}{Host redshift} & \multicolumn{3}{c}{GW skymap} & \multicolumn{2}{c}{ParSNIP} \\
        \colhead{} & \colhead{$z_{\rm host}$} & \colhead{$z_{\rm host}$ source} & \colhead{$d_L$ [Mpc]} & \colhead{2D CI} & \colhead{3D CI} & \colhead{Classification} & \colhead{Prob.}
    }
    \startdata
        AT 2023adwb & - & - & $1360 \pm 430$ & 0.136 & - & - & - \\
        AT 2023uho & - & - & $1250 \pm 552$ & 0.484 & - & - & - \\
        AT 2023adwp & - & - & $1274 \pm 433$ & 0.652 & - & - & - \\
        AT 2023aagj & 0.184 & Specz (this work) & $1478 \pm 425$ & 0.754 & 0.829 & Non-SN & 0.944 \\
        AT 2023adwt & - & - & $1429 \pm 413$ & 0.756 & - & - & - \\
        AT 2023uea & $0.195 \pm 0.084$ & Quaia SPz & $1459 \pm 426$ & 0.790 & 0.814 & Non-SN & 0.991 \\
        \hline
        \multicolumn{8}{c}{\textit{Transients without peak}} \\
        \hline
        AT 2023adio & $0.212 \pm 0.097$ & Quaia SPz & $1410 \pm 446$ & 0.242 & 0.363 & SNII & 0.901 \\
        AT 2023adwd & - & - & $1328 \pm 451$ & 0.285 & - & - & - \\
        AT 2023adwe & $1.528 \pm 0.671$ & LS photz & $1293 \pm 463$ & 0.285 & - & Non-SN & 0.927 \\
        AT 2023adwn & $0.349 \pm 0.078$ & LS photz & $875 \pm 486$ & 0.617 & 0.810 & Non-SN & 0.676 \\
        AT 2023adwq & $0.391 \pm 0.112$ & Quaia SPz & $1545 \pm 454$ & 0.683 & 0.784 & Non-SN & 0.968 \\
        AT 2023adfo & - & - & $1471 \pm 422$ & 0.729 & - & - & - \\
        \hline
        \multicolumn{8}{c}{\textit{Reddening transients}} \\
        \hline
        AT 2023adwc & 0.216 & NED specz & $1228 \pm 445$ & 0.143 & 0.080 & Non-SN & 0.996 \\
        AT 2023adwf & 0.180 & NED specz & $1148 \pm 417$ & 0.300 & 0.155 & Non-SN & 0.985 \\
        AT 2023adwj & - & - & $1345 \pm 457$ & 0.429 & - & - & - \\
        AT 2023uec & - & - & $1127 \pm 469$ & 0.459 & - & - & - \\
        AT 2023uos & - & - & $1448 \pm 458$ & 0.475 & - & - & - \\
        AT 2023adwl & - & - & $1051 \pm 441$ & 0.583 & - & - & - \\
        AT 2023unj & - & - & $1127 \pm 439$ & 0.642 & - & - & - \\
        AT 2023unl & 0.248 & Specz (this work) & $1535 \pm 444$ & 0.677 & 0.652 & - & - \\
        AT 2023adws & 0.178 & NED specz & $1451 \pm 419$ & 0.750 & 0.827 & Non-SN & 0.928 \\
        \hline
        \multicolumn{8}{c}{\textit{Excluded via spectra}} \\
        \hline
        AT 2023aden & - & - & $1154 \pm 451$ & 0.658 & - & - & - \\
        AT 2023uab & 0.128 & Specz (this work) & $919 \pm 479$ & 0.785 & 0.554 & SNIa & 0.814 \\
    \enddata
\end{deluxetable*}